\documentclass[structabstract]{aa}  

\usepackage{multirow} 
\usepackage{natbib}
\usepackage{amssymb,amsmath}
\usepackage{graphicx, subfigure}
\usepackage{longtable}

\usepackage{txfonts}
\usepackage[cmyk]{xcolor}

\newcommand{\PaperI}{Paper I }

\begin{document}

   \title{The wind of W Hya as seen by Herschel.}
\subtitle{II. The molecular envelope of W Hya}

   \author{T. Khouri
           \inst{1}\thanks{{\it Send offprint requests to T. Khouri}\newline \email{theokhouri@gmail.com}}, A. de Koter \inst{1,2}, L. Decin\inst{1,2}, L. B. F. M. Waters \inst{1,3},
           M. Maercker \inst{4,5}, R. Lombaert \inst{2}, J. Alcolea\inst{6}, J. A. D. L. Blommaert\inst{2,7}, V. Bujarrabal \inst{8}, M. A. T. Groenewegen \inst{9}, K. Justtanont \inst{4},
F. Kerschbaum \inst{10}, M. Matsuura\inst{11}, K. M. Menten \inst{12}, H. Olofsson \inst{4}, P. Planesas \inst{6}, P. Royer \inst{2}, M. R. Schmidt \inst{13}, R. Szczerba \inst{13},
D. Teyssier \inst{14}, J. Yates \inst{11}}
\institute{Astronomical Institute ÒAnton PannekoekÓ, University of Amsterdam, PO Box 94249, 1090 GE Amsterdam, The Netherlands %1
         \and
            Instituut voor Sterrenkunde, KU Leuven, Celestijnenlaan 200D B-2401, 3001 Leuven, Belgium %2
            \and
            SRON Netherlands Institute for Space Research, Sorbonnelaan 2, 3584 CA Utrecht, The Netherlands %3
            \and
	Department of Earth and Space Sciences, Chalmers University of Technology, Onsala Space Observatory, SE-439 92 Onsala, Sweden%4
	            \and
	            Argelander Institute f\"ur Astronomie, Universit\"at Bonn, Auf dem H\"ugel 71, 53121 Bonn, Germany %5
	            \and
            Observatorio Astron\'omico Nacional (IGN), Alfonso XII N$^\circ$3, E-28014 Madrid, Spain %6
            \and
            Department of Physics and Astrophysics, Vrije Universiteit Brussel, Pleinlaan 2, 1050 Brussels, Belgium %7
            \and
            Observatorio Astron\'omico Nacional (OAN-IGN), Apartado 112, E-28803 Alcal\'a de Henares, Spain %8
            \and
            Koninklijke Sterrenwacht van Belgi\"e, Ringlaan 3, B--1180 Brussel, Belgium %9
            \and
            University of Vienna, Department of Astrophysics, T\"urkenschanzstra\ss e 17, 1180 Wien, Austria %10
            \and
            Dept. of Physics \& Astronomy, University College London, Gower St, London WC1E 6BT, UK %11
	\and
	            Max-Planck-Institut f\"ur Radioastronomie, Auf dem H\"ugel 69, 53121 Bonn, Germany %12
            \and
            N. Copernicus Astronomical Center, Rabia\'nska 8, 87-100 Toru\'n, Poland %13
            \and
            European Space Astronomy Centre, Urb. Villafranca del Castillo, PO Box 50727, E-28080 Madrid, Spain}%14
 \titlerunning{The molecular wind of W Hya}
\authorrunning{Khouri et al.}
  \abstract
{The evolution of low- and intermediate-mass stars on the asymptotic giant branch (AGB) is mainly controlled by the rate at which these stars lose mass in a stellar wind.  Understanding the driving mechanism and strength of the stellar winds of AGB stars and the processes enriching their surfaces with products of nucleosynthesis are paramount to constraining AGB evolution and predicting the chemical evolution of galaxies.}
{In a previous paper we have constrained the structure of the outflowing envelope of W Hya using spectral lines of the $^{12}$CO molecule.  Here we broaden this study by including an extensive set of H$_{2}$O
and $^{28}$SiO lines.  It is the first time such a comprehensive study is performed for this source.  The oxygen isotopic ratios and the $^{28}$SiO abundance profile can be connected to the initial stellar mass and to
crucial aspects of dust formation at the base of the stellar wind, respectively.}
   {We model the molecular emission observed by the three instruments on board Herschel Space Observatory using a state-of-the-art molecular excitation
   and radiative transfer code. We also account for the dust radiation field in our calculations.}
{We find an H$_2$O ortho-to-para ratio of 2.5\,$^{+2.5}_{-1.0}$, consistent with what is expected for an AGB wind. The O$^{16}$/O$^{17}$ ratio indicates that W Hya has an initial
mass of about 1.5 M$_\odot$. Although the ortho- and para-H$_{2}$O lines observed by HIFI appear to trace gas of slightly different physical properties, we find that a turbulence
velocity of $0.7\pm0.1$ km s$^{-1}$ fits the HIFI lines of
both spin isomers and those of $^{28}$SiO well.}
{The modelling of H$_{2}$O and $^{28}$SiO confirms the properties of the envelope model of W Hya, as derived from $^{12}$CO lines, and allows us to constrain the turbulence velocity. 
The ortho- and para-H$_2^{16}$O and $^{28}$SiO abundances relative to H$_{2}$ are $(6^{+3}_{-2}) \times 10^{-4}$, $(3^{+2}_{-1}) \times 10^{-4}$, and $(3.3\pm 0.8)\times 10^{-5}$,
respectively, in agreement with expectations for oxygen-rich AGB outflows.  Assuming a solar silicon-to-carbon ratio, the $^{28}$SiO line emission model is consistent with about one-third
of the silicon atoms being locked up in dust particles.}


\maketitle

\section{Introduction}

Stars of low- and intermediate-mass ($\sim$\,0.8 to 8 M$_\odot$) populate the asymptotic giant branch (AGB) at the end of their lives.
These luminous, extended, and cool objects develop a strong mass loss that controls their evolution from its onset.  
The driving of the wind is thought to be the result of a combination of pulsations and radiation pressure
on dust grains \citep[see e.g.][]{Habing2003}.
For carbon-rich AGB stars, this scenario is able to reproduce the observed mass-loss rates as the dust species
that form in these environments are opaque enough and can exist close enough to the star to acquire momentum by
absorbing infrared photons \citep{Winters2000}.
However, for oxygen-rich AGB stars the situation seems more complex. 
The dust species found in these stars through infrared spectroscopy that would be able to drive the
wind by absorption of stellar photons cannot exist close enough to the star to be important for initiating the wind \citep{Woitke2006}.
A possible alternative is that the wind is driven through scattering
of photons on large ($\sim 0.3\,\mu$m) translucent dust grains \citep[][]{Hofner2008}.
Although this alternative scenario seems plausible and many dust species have been identified in oxygen-rich objects,
it is still unclear which of these are actually responsible for driving the outflow \cite[e.g.][]{Bladh2012}.

Characterizing both the physical structure and molecular and solid state composition of the outflow are crucial to
understanding the physics underlying the wind driving. An important step forward in this understanding is 
the combined analysis of space and ground-based observations of multiple molecular species, covering a range
of rotational excitation states of the ground vibrational level as wide as possible.  Such data allow the determination of the flow properties
from the onset of the wind close to the photosphere, to the outer regions, where molecules
are eventually photodissociated by the interstellar radiation field.  We have embarked on such an analysis for W\,Hya,
a close-by oxygen-rich AGB star, observed in detail using the Herschel Space Observatory \citep[hereafter {\em Herschel};][]
{Pilbratt2010}. In addition to the $^{12}$CO lines, the observations reveal a rich H$_2^{16}$O spectrum with 
over 150 observed lines, and a broad $^{28}$SiO ladder, ranging from 137 to more than 2000 K in upper level excitation energy.

In \citet[][henceforth Paper I]{Khouri2014}, we focussed on the analysis of the $^{12}$CO ladder of rotational levels up
to $J_{\rm up}= 30$.  We used these carbon monoxide lines to determine the temperature and velocity 
structure of the wind, as well as the mass-loss rate. In the study presented here we focus on an analysis of lines from H$_2$O,
including its isotopologues, and $^{28}$SiO. The modelling of the H$_2$O isotopologues allows us to constrain
the $^{17}$O/$^{16}$O and $^{18}$O/$^{16}$O ratios.  The analysis of silicon in both the gas phase and solid phase allows us to assess the overall budget of this 
element and the condensation fraction of this species.  This is a crucial issue for understanding the role of silicates
for the wind driving mechanism. We also account for the contribution of the dust thermal emission to
the radiation field when modelling the molecular emission.
The main components of these grains are 
aluminum-oxides and silicates \citep[e.g.][]{Sharp1990}.

Determining the ortho- and para-H$_2$O abundances is important for understanding the chemistry in
the outermost layers of the star and the innermost regions of the envelope, where shocks can be important \citep{Cherchneff2006}.
For an AGB star with carbon-to-oxygen ratio of 0.75, \cite{Cherchneff2006} finds the abundance ratio between $^{12}$CO and H$_2^{16}$O
to be $\sim 12$ and $\sim 1.6$ for thermal equilibrium and non-equilibrium chemistry respectively.
The H$_2$O ortho-to-para ratio is expected to be 3:1, reflecting the ratio of the statistical
weights between the species, if the molecules are formed in a high-temperature (T\,$\gg 30$ K)
and under local-thermodynamical equilibrium \citep{Decin2010}.
Studies of the H$_2^{16}$O emission from W\,Hya have found a low ortho-to-para ratio of around 1 \citep[e.g.][]{Barlow1996,Justtanont2005,Zubko2000}.
The uncertainties on these obtained ortho-to-para ratios are, however, at least of a factor of two.

The isotopic composition of the outflowing material contains much-needed information on the dredge-up processes
that are an important part of giant branch and AGB evolution \citep[e.g.][]{Landre1990,ElEid1994,Stoesz2003,Charbonnel2010,Palmerini2011}.
Dredge-ups enrich the surface with
isotopic species when convective streams in the star reach down to regions where the composition has been modified
because of thermonuclear burning \citep[see e.g.][]{Iben1983,Iben1975}.  The characteristics of the
isotopic enrichment are expected to vary significantly over the evolution of low- and intermediate-mass stars and
are found to be especially sensitive to stellar mass \citep[][]{Boothroyd1994} --  the most important stellar property, 
which for AGB stars is notoriously difficult to constrain.   Unfortunately, the dredge-up process in AGB stars cannot 
yet be modelled from first principles \citep[][]{Habing2003}, mainly because of the complex (and poorly understood) physics 
of convective and non-convective mixing \citep[see e.g.][]{Busso1999,Karakas2010}. 

In Section \ref{sec:info} we provide general information on W\,Hya, the $^{12}$CO model of Paper I, and the dataset that is used to 
constrain the H$_2^{16}$O and $^{28}$SiO properties.  In Section \ref{sec:comp_mol}, we discuss observed line shapes and
provide details on the treatment of H$_2^{16}$O and $^{28}$SiO in our models, and how this treatment relates to that of $^{12}$CO.
Section \ref{sec:WHya_H2O} is devoted to presenting our model for H$_2^{16}$O and for the lower abundance isotopologues. The model for $^{28}$SiO is covered 
in Section \ref{sec:SiO}. We discuss the results in Section \ref{sec:disc} and we end with a summary.

\section{Basis information, dataset, and model assumptions}
\label{sec:info}

\subsection{W\,Hya}

An overview of the literature discussing the stellar properties of W\,Hya, and the uncertainties in these properties, is presented in Paper I. We refer
to this paper for details.  We adopt a distance to the star of 78 pc \citep{Knapp2003}, implying a luminosity of 5400 $L_{\odot}$.  Assuming that W\,Hya
radiates as a black-body, our $^{12}$CO analysis
is consistent with an effective temperature of 2500\,K, which leads to a stellar radius, R$_\star$, of $2.93 \times 10^{15}$\,cm or 1.96 AU. Whenever we
provide the radial distance in the wind of W\,Hya in units of $R_{\star}$, this is the value we refer to.  The $^{12}$CO analysis presented in Paper I leads to a 
mass-loss rate of
$1.3 \times 10^{-7}$ $M_\odot$\,yr$^{-1}$, consistent with the findings of \cite{Justtanont2005}.  Interestingly, on a scale that is larger than
the $^{12}$CO envelope, images of cold dust emission suggest W Hya had a substantially larger mass loss some $10^{3}$--$10^{5}$ years ago
\citep{Cox2012,Hawkins1990}. In this work, we focus on the present-day mass-loss rate by modelling the gas-phase $^{28}$SiO and H$_2$O
emission, which probe the last 200 years at most. Variations in the mass-loss rate seen on larger timescales will be addressed in a future study.

W\,Hya features prominent rotational H$_2^{16}$O emission, first reported by \cite{Neufeld1996} and \cite{Barlow1996}, using data obtained with the {\em Infrared Space
Observatory} \citep[ISO;][]{Kessler1996}.  Additional data were obtained by \cite{Justtanont2005}, using {\em Odin} \citep{Nordh2003}, and
by \cite{Harwit2002}, using SWAS \citep{Melnick2000}.  The analysis by the authors mentioned above and by others 
\citep{Zubko2000,Maercker2008, Maercker2009} point to a quite high H$_2^{16}$O abundance relative to H$_{2}$, ranging from $10^{-4}$ to a few 
times $10^{-3}$.  The ortho-to-para H$_{2}$O ratio reported by most studies is usually in-between 1 and 1.5, which is significantly lower than 
the expected value of three, for H$_2$O formed at high temperatures (T\,$\gg 30$ K) and in local-thermodynamical equilibrium.

$^{28}$SiO lines of the vibrational excited $v = 1$ and $v = 2$ state show maser emission and have been intensively 
studied \citep[see e.g.][]{Imai2010,Vlemmings2011}.  Ground vibrational lines do not suffer from strong amplification and are better probes
of the silicon abundance.  Studies of this molecule \citep[e.g.][]{GonzalezDelgado2003,Bieging2000,Bujarrabal1986} suggest $^{28}$SiO is
depleted from the gas, probably due to the formation of silicate grains.
Spatially resolved data of the $v=0, J=2-1$ line have been presented by \cite{Lucas1992}. The authors
determined the half-intensity radius of this transition to be 0.9 arcseconds. From model calculations,
\cite{GonzalezDelgado2003} conclude that the half-intensity radius determined by \cite{Lucas1992} is approximately three times
smaller than the radius where the $^{28}$SiO abundance has decreased to 37\% of its initial value.

The dust envelope of W\,Hya has been imaged by \cite{Norris2012} using aperture-masking polarimetric interferometry.
They discovered a close-in shell of large ($\sim$\,0.3 $\mu$m) and translucent grains which might be responsible for
driving the outflow through scattering \citep{Hofner2008}. Unfortunately, the composition of the observed grains could,
not be unambiguously identified. W\,Hya has also been observed in the near-infrared using MIDI/VLTI \citep{Zhao-Geisler2011}.
This revealed that the 
silicate dust emission must come from an envelope with an inner radius of  50 AU (or 28 $R_{\star}$). 
\cite{Zhao-Geisler2011} also argue that Al$_2$O$_3$ grains and H$_{2}$O molecules close to the star
are responsible for the observed increase in diameter at wavelengths longer than $10\ \mu$m.

\subsection{Dataset}
\label{sec:dataset}

W\,Hya was observed by all three instruments onboard {\em Herschel} in the context of the guaranteed-time key programs HIFISTARS \citep{Menten2010}
and MESS \citep{Groenewegen2011}. 
These are the Heterodyne Instrument for the Far Infrared \citep[HIFI;][]{deGraauw2010},
the Spectral and Photometric Imaging Receiver Fourier-Transform Spectrometer
\citep[SPIRE;][]{Griffin2010}, and the Photodetector Array Camera and Spectrometer \cite[PACS;][]{Poglitsch2010}. 
The data reduction procedure of the PACS and SPIRE data is outlined in \PaperI; that of HIFI is presented by 
\cite{Justtanont2012}.

We applied two methods to identify the H$_2^{16}$O and $^{28}$SiO lines. The $^{28}$SiO lines were identified in very much
the same way as the $^{12}$CO lines presented in Paper I: we inspected the spectra at the wavelengths of the $^{28}$SiO transitions, 
identified the lines and measured their fluxes. H$_2^{16}$O, however, has a plethora of transitions that, moreover, are not regularly 
spaced in frequency, as are the ones of $^{12}$CO and $^{28}$SiO.  Our approach in this case was to calculate spectra with 
different values for the H$_2^{16}$O-envelope parameters and compare these to the observations. Transitions 
that were predicted to stand out from the noise were identified and extracted.

In order to extract the integrated fluxes from the PACS spectrum, we fitted Gaussians to the identified transitions using 
version 11.0.1 of {\em Herschel} interactive processing environment (HIPE\footnote{HIPE is available for download at http://herschel.esac.esa.int/hipe/}).  For extracting transitions measured by SPIRE, we 
used the script {\it Spectrometer Line Fitting} available in HIPE. The script simultaneously fits a power law to the continuum 
and a cardinal sine function to the lines in the unapodized SPIRE spectrum.
The spectral resolution of both instruments is smaller than twice the terminal velocity of W\,Hya ($\upsilon_{\infty}$ = 7.5 kms$^{-1}$;
see Table~\ref{tab:grid_par}). Therefore, 
what may appear to be a single observed line might be a 
blend of two or more transitions. We removed such blended lines from our analysis by flagging them as blends
whenever two or more transitions were predicted to be formed closer together than the native full width at 
half-maximum (FWHM) of the instrument, or when the FWHM of the fitted Gaussian was 20\%
or more larger than the expected FWHM for single lines. 
For PACS, a total of 50 ortho-H$_2^{16}$O, 24 para-H$_2^{16}$O, and only 1 $^{28}$SiO unblended transitions were extracted.
The $^{28}$SiO transitions that lie in the spectral region observed by PACS are high excitation lines ($J_{\rm up} > 35$). These are
weak compared to the much stronger H$_2^{16}$O lines and are below the detection limit of the PACS spectrum for $J_{\rm up} > 38$.
For SPIRE, 15 ortho-H$_2^{16}$O, 20 para-H$_2^{16}$O and 24 $^{28}$SiO lines were extracted and not flagged as 
blends. Properties of all H$_2^{16}$O transitions measured by PACS and those of $^{28}$SiO measured by all instruments are given in 
Appendix \ref{sec:AppendixB}.

The transitions observed with HIFI are spectrally resolved.
This gives valuable information on the velocity structure of the flow.
Because of the high spectral resolution the lines are easily identified and their total fluxes can be measured accurately.
In Table~\ref{tab:obs_HIFI}, we list the integrated main beam brightness temperatures and the excitation energy
of the upper level for the observed transitions.
HIFI detected ten ortho-H$_2^{16}$O transitions, two of which are clearly masering, six para-H$_2^{16}$O transitions, 
one of which is clearly masering, five $^{28}$SiO transitions,
three of which are from vibrationally excited states and appear to be masering,
two transitions of ortho-H$_2^{17}$O and $^{29}$SiO, and one transition each of ortho-H$_2^{18}$O,  para-H$_2^{17}$O, para-H$_2^{18}$O and $^{30}$SiO.
We have not included the rarer isotopologues of SiO, $^{29}$SiO and $^{30}$SiO, in our model calculations.
The vibrationally excited lines from $^{28}$SiO probe high temperature gas (T\,$\sim 2000$\,K) which is very close to the star and were not included in our analysis.
The upper level excitation energies of the $^{28}$SiO ground-vibrational lines range from 137 to 1462 K and these transitions probe the gas temperatures in which silicates
are expected to condense, T\,$\sim1000$\,K \citep[e.g.][]{Gail1999}.

\subsubsection{H$_2^{16}$O masers}

The already challenging task of modelling H$_2^{16}$O emission is further complicated by the fact that this molecule 
has the predisposition to produce maser emission.  \cite{Matsuura2013}
found that for the oxygen-rich supergiant VY\,CMa 70\% of the H$_2^{16}$O lines in the PACS and SPIRE spectra are affected 
by population inversions.
In our line emission code we artificially suppress strong stimulated emission and we have therefore excluded from the present analysis lines
that show masering in at least one of our models. In the PACS data, four transitions were excluded. Those
are the ones at 133.549, 159.051 and 174.626 $\mu$m of ortho-H$_2^{16}$O and 138.528 $\mu$m of para-H$_2^{16}$O.
The intensity of the ortho-H$_2^{16}$O transition at 73.415 was under-predicted by more than a factor of three by all models.
Although this line is not masering in any of our models, the discrepancies are so large that we suspect missing physics to be responsible.
We have also excluded this line from the analysis.
In four of the remaining PACS lines, we do find modest population inversions for some of the models. These, however,
do not appear to have a strong impact on the strength of the lines and the transitions were not excluded.
For the SPIRE lines, the case is worse. About 70 \% of the lines -- in agreement with the results reported for VY\,CMA by \cite{Matsuura2013} -- 
present population inversions. The intensity of most of these lines was strongly under-predicted by our models. This is a much higher fraction
than the less than 10\% seen for PACS. To be on the safe side, we decided not to include SPIRE data at all in our calculations for H$_2^{16}$O.
As a result of this mismatch a comprehensive identification of the H$_2^{16}$O lines in the SPIRE spectrum was not possible and we do not present extracted line fluxes
for H$_2^{16}$O lines obtained by this instrument.

\subsubsection{Comparison between the total line fluxes measured by HIFI, SPIRE and PACS}

Some of the lines were observed by more than one instrument, either by HIFI and SPIRE or by HIFI and PACS.
In Table~\ref{tab:obs_comp} we present a comparison of the observed total line fluxes. The antennae temperatures 
observed by HIFI were converted to flux units assuming that the emission is not spatially resolved by the telescope.
We find that the fluxes of H$_2^{16}$O lines as detected by HIFI are systematically lower than those by both SPIRE 
and PACS.

SPIRE lines are typically 25\% stronger than those of HIFI. This difference may be due to uncertainties
in the flux calibration of both instruments and in the baseline fitting of the spectra. Small blends and pointing errors can also 
account for some of the observed differences. A flux loss of less than 10\% is expected from the pointing errors in the HIFI 
observations. The different shapes of the response function of the instruments may also have a significant effect.

The three lines measured jointly by PACS and HIFI are more than 2--3 times stronger in the PACS traces than
in the HIFI spectra. The cross-calibration problem of PACS and HIFI is known, but is not yet resolved.
It is often reported as a flux mismatch between the $^{12}$CO lines observed by both instruments
(Puga et al. in prep.).
For W\,Hya we find that the mismatch between predicted and PACS line fluxes correlates with wavelength, becoming
larger for longer wavelengths.  At the long wavelength end of the PACS spectral range (which is where PACS and
HIFI spectra overlap) the discrepancy strongly increases.
Since the HIFI observations are much less susceptible to line blending and the HIFI line fluxes
in this overlap region agree much better with the trends predicted by our models than the line fluxes measured by PACS,
we decided to exclude these PACS lines from our analysis, and to use the HIFI lines instead.

\subsection{Modelling strategy and $^{12}$CO model}

Here, we focus on modelling the emission of ortho-H$_2^{16}$O, para-H$_2^{16}$O, ortho-H$_2^{18}$O, 
para-H$_2^{18}$O, ortho-H$_2^{17}$O, para-H$_2^{17}$O and $^{28}$SiO. The molecular data used in our calculations are described
in Appendix \ref{sec:AppendixA}. Our model is based on the envelope structure and
dust model obtained in Paper I, which was obtained from the analysis of $^{12}$CO lines, observed with {\it Herschel}, APEX, SMT and SEST, and of the dust emission.
The $^{12}$CO transitions modelled in Paper I probe a large range in excitation temperature and, therefore, the full extent of the outflowing $^{12}$CO envelope,
from the regions close to the star where the wind is accelerated, to the outer regions where $^{12}$CO is photo-dissociated.
The dust properties were constrained by modelling the thermal emission spectrum observed
by ISO with the continuum radiative-transfer code MCMax \citep{Min2009}.  The composition and radial distribution of
the dust are used as inputs to the modelling of the molecular species, for which we employ GASTRoNOoM
\citep{Decin2006,Decin2010a}. The coupling between dust and gas is treated as described by \cite{Lombaert2013}.
The main parameters of the $^{12}$CO model are listed in Table~\ref{tab:grid_par}. 

\begin{table}[t!]
\caption{Model parameters of W\,Hya as derived in Paper I.}
\centering
\label{tab:grid_par}
\begin{tabular}{ c | c }
\hline\\[-8pt]
Parameter & Best Fit \\[1pt]
\hline\\[-8pt]
$dM/dt$ [$10^{-7}$ M$_{\odot}$\,yr$^{-1}$] & $1.3 \pm 0.5$\\
$\upsilon_{\infty}$ [km\,s$^{-1}$] & $7.5 \pm 0.5$\\
$\epsilon$ & $0.65 \pm 0.05$\\
$\upsilon_{\rm turb}$ [km\,s$^{-1}$] & $1.4 \pm 1.0$ \\
$f_{\rm CO}$ &$ 4 \times 10^{-4}$ \\
R$_{1/2}$ & 0.4\\
T$_*$ [K] & 2500 \\
\hline
\end{tabular}
\tablefoot{$\epsilon$ is the exponent of the temperature power law and R$_{1/2}$ is the radius where the CO abundance has decreased by half.}
\end{table}

In Paper I, we found that the $^{12}$CO envelope has to be smaller than predicted for our model to better fit the high-
($J_{\rm up} \ge 6$) and low-excitation ($J_{\rm up} < 6$) $^{12}$CO transitions simultaneously.
The size of the $^{12}$CO envelope is set in our model by the parameter R$_{1/2}$ \citep{Mamon1988},
which represents the radius at which the $^{12}$CO abundance has decreased by half.
We note that this discrepancy between model and observations 
in the outer wind is not expected to affect the modelled H$_2^{16}$O or $^{28}$SiO lines, since these molecules occupy a much smaller 
part of the envelope than $^{12}$CO, see Fig. \ref{fig:regions_probed}.

Furthermore, to properly represent the PACS and HIFI transitions excited in the inner part of the wind, our best model
requires a value of 5.0 for the exponent of the $\beta$-type velocity law (see Equation \ref{eq:beta-law}),
\begin{equation}
\label{eq:beta-law}
\upsilon(r) = \upsilon_{\circ} + (\upsilon_{\infty} - \upsilon_{\circ}) \left(1 - \frac{r_{\circ}}{r} \right)^{\beta}.
\end{equation}
This corresponds to a slow acceleration 
of the flow in this part of the envelope.  However, this high value of $\beta$ underpredicts the width of the $^{12}$CO 
$J$ = 6\,--\,5 transition, while predicting well the width of lower excitation transitions ($J_{\rm up}<6$). This indicates a more rapid acceleration
between the formation region of transitions $J$ = 10\,--\,9 and $J$ = 6\,--\,5 than considered in our  $\beta = 5.0$ model. This could be
due to the addition of extra opacity in the wind at distances beyond where the $J$ = 10\,--\,9 transition forms, which is roughly inside 50 R$_\star$.
As the H$_2^{16}$O and $^{28}$SiO envelopes are smaller 
than the excitation region of the $^{12}$CO $J_{\rm up} = 6$ level, emission from these two molecules comes from a wind region that is well 
described by a $\beta = 5.0$ velocity law (see also Fig.~\ref{fig:regions_probed}).

Our dust model is motivated by the work of \cite{Justtanont2004,Justtanont2005} and adopts a dust 
mass-loss rate of $2.8 \times 10^{-10}$ M$_{\odot}$\,yr$^{-1}$. Astronomical silicates, amorphous aluminum oxide 
(Al$_2$O$_3$), and magnesium-iron oxide (MgFeO) account for 58, 34, and 8\% of the dust mass, respectively.
The optical constants for astronomical silicates are from \cite{Justtanont1992}, those for amorphous aluminum oxide and 
magnesium-iron oxide were retrieved from the University of Jena database and are from the works of \cite{Begemann1997}
and \cite{Henning1995}.

\begin{table}[t!]
\caption{H$_{2}$O and $^{28}$SiO transitions observed by HIFI for W Hya.}
\label{tab:obs_HIFI}
\centering
\begin{tabular}{ l | r r@{.}l l@{$\pm$}l   } 
\hline\\[-8pt]    
Transitions & $\nu_0$ [GHz] & \multicolumn{2}{c}{$E$ [K]} & \multicolumn{2}{r}{$\int$  $T_{\rm MB} \,dv$ [K km\,s$^{-1}$]} \\[2pt]
\hline\\[-8pt]
o-H$_2^{16}$O$_{\nu = 0,\,5_{3,2}-5_{2,3}}$ 			& 1867.749 	& 732&0 	& $ 5.3$ & $ 0.6 $ \\[1pt]
o-H$_2^{16}$O$_{\nu = 0,\,7_{3,4}-7_{2,5}}$ 			& 1797.159 	& 1212&0 	& $ 2.3$ & $ 0.3 $ \\[1pt]
o-H$_2^{16}$O$_{\nu_2 = 1,\,2_{1,2}-1_{0,1}}$ 			& 1753.914 	& 2412&9 	& $2.2$ & $ 0.6$ \\[1pt]
o-H$_2^{16}$O$_{\nu = 0,\,3_{0,3}-2_{1,2}}$ 			& 1716.769  	& 196&8 	& $ 24.9$ & $ 0.3 $ \\[1pt]
o-H$_2^{16}$O$_{\nu = 0,\,3_{2,1}-3_{1,2}}$ 			& 1162.911 	& 305&2 	& $8.1$ & $ 0.3$ \\[1pt]
o-H$_2^{16}$O$_{\nu = 0,\,3_{1,2}-2_{2,1}}$ 			& 1153.127 	& 249&4 	& $21.1$ & $ 0.4$ \\[1pt]
o-H$_2^{16}$O$_{\nu = 0,\,3_{1,2}-3_{0,3}}$ 			& 1097.365 	& 249&4 	& $8.0$ & $ 0.2$ \\[1pt]
o-H$_2^{16}$O$_{\nu_2 = 1,\,1_{1,0}-1_{0,1}}$ 			& 658.007 	& 2360&3 	& $28.9$ & $ 0.1$ \\[1pt]
o-H$_2^{16}$O$_{\nu = 0,\,5_{3,2}-4_{4,1}}$ 			& 620.701 	& 732&1 	& $8.4$ & $ 0.1$ \\[1pt]
o-H$_2^{16}$O$_{\nu = 0,\,1_{1,0}-1_{0,1}}$ 			& 556.936 	& 61&0 	& $5.4$ & $ 0.1$ \\[1pt]
o-H$_{2}^{\ 17}$O$_{\nu = 0,\,3_{0,3}-2_{1,2}}$ 	& 1718.119 	& 196&4  	& $ 1.0$ & $ 0.2$ \\[1pt]
o-H$_2^{\ 17}$O$_{\nu = 0,\,3_{1,2}-3_{0,3}}$ 	& 1096.414 	& 249&1 	& $0.5$ & $ 0.1$ \\[1pt]
o-H$_2^{\ 18}$O$_{\nu = 0,\,3_{1,2}-3_{0,3}}$ 	& 1095.627 	& 248&7 	& $1.5$ & $ 0.1$ \\[1pt]
p-H$_2^{16}$O$_{\nu = 0,\,6_{3,3}-6_{2,4}}$ 			& 1762.043 	& 951&8 	& $2.4$ & $ 0.6$ \\[1pt]
p-H$_2^{16}$O$_{\nu = 0,\,4_{2,2}-4_{1,3}}$ 			& 1207.639 	& 454&3 	& $ 3.3$ & $ 0.5$ \\[1pt]
p-H$_2^{16}$O$_{\nu = 0,\,1_{1,1}-0_{0,0}}$ 			& 1113.343 	& 53&4 	& $15.4$ & $ 0.1$ \\[1pt]
p-H$_2^{16}$O$_{\nu = 0,\,2_{0,1}-1_{1,1}}$ 			& 987.927 	& 100&8 	& $18.6$ & $ 0.2$ \\[1pt]
p-H$_2^{16}$O$_{\nu = 0,\,5_{2,4}-4_{3,1}}$ 			& 970.315 	& 598&8 	& $20.9$ & $ 0.1$ \\[1pt]
p-H$_2^{16}$O$_{\nu = 0,\,2_{1,1}-2_{0,2}}$ 			& 752.033 	& 136&9 	& $7.0$ & $ 0.1$ \\[1pt]
p-H$_2^{\ 17}$O$_{\nu = 0,\,1_{1,1}-0_{0,0}}$ 	& 1107.167 	& 53&1 	& $0.5$ & $ 0.1$ \\[1pt]
p-H$_2^{\ 18}$O$_{\nu = 0,\,1_{1,1}-0_{0,0}}$ 	& 1101.698 	& 52&9 	& $1.5$ & $ 0.1$ \\[1pt]
$^{28}$SiO$_{\nu = 1,\,J=23-22}$				& 990.355 	& 2339&9 	& $0.4$ & $ 0.1$ \\[1pt]
$^{28}$SiO$_{\nu = 0,\,J=16-15}$				& 694.275 	& 283&3 	& $2.9$ & $ 0.1$ \\[1pt]
$^{28}$SiO$_{\nu = 1,\,J=15-14}$				& 646.429 	& 2017&4 	& $0.2$ & $ 0.04$ \\[1pt]
$^{28}$SiO$_{\nu = 0,\,J=14-13}$				& 607.599 	& 218&8 	& $2.8$ & $ 0.1$ \\[1pt]
$^{28}$SiO$_{\nu = 1,\,J=13-12}$				& 560.326 	& 1957&4 	& $0.3$ & $ 0.1$ \\[1pt]
\hline
\end{tabular}
\end{table}

\begin{table*}
\centering
\caption{Observed total line fluxes of H$_2^{16}$O and $^{28}$SiO transitions observed by two different {\em Herschel} instruments, HIFI and SPIRE or PACS.
              The final column lists the flux of either SPIRE or PACS, relative to HIFI.}
\label{tab:obs_comp}
\begin{tabular}{ c | r r r l l l r  } 
\hline\\[-8pt]
Transition & $\nu_0$ & $\lambda$ & E$_{\rm up}$ & Flux HIFI & Flux SPIRE & Flux PACS &  F(Other)/F(HIFI) \\
&  [GHz] & [$\mu$m] & [K] & [W\,m$^{-2}$] & [W\,m$^{-2}$]  & [W\,m$^{-2}$]  &  \\[2pt]
\hline
\hline\\[-8pt]
ortho-H$_2^{16}$O & & & & & & & \\[2pt]
1$_{1,0}-1_{0,1}$ & 556.936 & 538.29 & 61 & 3.8 $\times 10^{-17}$ & 5.5 $\times 10^{-17}$ & - & 1.45 \\
5$_{3,2}-4_{4,1}$ & 620.701 & 482.99 & 732 & 6.5 $\times 10^{-17}$ & 7.4 $\times 10^{-17}$ & - & 1.14 \\
v2=1,1$_{1,0}-1_{0,1}$ & 658.007 & 455.61 & 2360 & 2.4 $\times 10^{-16}$ & 3.0 $\times 10^{-16}$ & - & 1.25 \\
3$_{1,2}-3_{0,3}$ & 1097.365 & 273.19 & 249 & 1.1 $\times 10^{-16}$ & 1.3 $\times 10^{-16}$ & - & 1.18 \\
3$_{1,2}-2_{2,1}$ & 1153.127 & 259.98 & 249 & 3.0 $\times 10^{-16}$  & 3.7 $\times 10^{-16}$ & - & 1.23 \\
3$_{2,1}-3_{1,2}$ & 1162.911 & 257.79 & 305 & 1.2 $\times 10^{-16}$ & 1.8 $\times 10^{-16}$ & - & 1.50 \\
3$_{0,3}-2_{1,2}$ & 1716.769 & 174.63 & 197 & 5.4 $\times 10^{-16}$ & - &  1.3 $\times 10^{-15}$ & 2.41 \\
5$_{3,2}-5_{2,3}$ &  1867.749 & 160.51 & 732 & 1.2 $\times 10^{-16}$  &  - & 2.8 $\times 10^{-16}$ & 2.33 \\[2pt]
para-H$_2^{16}$O & & & & & & & \\[2pt]
2$_{1,1}-2_{0,2}$ & 752.033 & 398.64 & 137 & 6.6 $\times 10^{-17}$ & 8.1 $\times 10^{-17}$ & - & 1.23 \\
5$_{2,4}-4_{3,1}$ & 970.315 & 308.96 & 599 & 2.5 $\times 10^{-16}$ & 3.6 $\times 10^{-16}$ & - & 1.44 \\
2$_{0,2}-1_{1,1}$ & 987.927 & 303.46 & 101 & 2.3 $\times 10^{-16}$ & 2.9 $\times 10^{-16}$ & - & 1.26 \\
1$_{1,1}-0_{0,0}$ & 1113.343 & 269.27 & 53 & 2.2 $\times 10^{-16}$ & 2.2 $\times 10^{-16}$ & - & 1.00 \\
4$_{2,2}-4_{1,3}$ & 1207.639 & 248.25 & 454 & 4.9 $\times 10^{-17}$ & 6.7 $\times 10^{-17}$ & - &  1.21 \\
6$_{3,3}-6_{2,4}$ & 1762.043 & 170.14 & 952 &  5.3 $\times 10^{-17}$ &  - & 1.7 $\times 10^{-16}$ & 3.21 \\[2pt]
$^{28}$SiO & & & & & & & \\[2pt]
$J$=14-13 & 607.608 & 493.40 & 218.8 & 2.1 $\times 10^{-17}$ & 2.0 $\times 10^{-17}$ & - & 0.95 \\
$J$=16-15 & 694.294 & 431.79 & 283.3 & 2.5 $\times 10^{-17}$ & 2.4 $\times 10^{-17}$ & - & 0.96 \\
\hline
\end{tabular}
\end{table*}

\section{Comparisons between $^{12}$CO, H$_2^{16}$O and $^{28}$SiO}
\label{sec:comp_mol}

\subsection{Photodissociation radii}
\label{sec:photodis}

$^{12}$CO, $^{28}$SiO, and H$_{2}^{16}$O are formed in the atmosphere of the AGB star and the fate of these molecules
is to become dissociated in the outer envelope by  interstellar UV photons, causing their abundances to decrease sharply.
The radius where dissociation sets in is different for each molecule and depends on molecular and 
circumstellar parameters and on the spectral shape of the interstellar radiation field.

Dust condensation might play a role in shaping the abundance profile of a molecule if the 
conditions for condensation are met before dissociation sets in.
For W\,Hya we expect $^{28}$SiO to condense and form silicate grains. $^{12}$CO 
and H$_2^{16}$O abundances should not be affected by such depletion and we assume their
abundances to be constant up
to the point where dissociation starts. For $^{28}$SiO
we consider different abundance profiles, mimicking depletion due to silicate 
formation (see Sect.~\ref{sec:SiO}).

Dissociation radii are usually poorly constrained.
In our $^{12}$CO model presented in Paper I, we find that for W\,Hya the $^{12}$CO envelope is likely smaller than expected for a standard
interstellar radiation field. Our results point to an envelope with $^{12}$CO being fully dissociated at roughly 800 $R_{\star}$.
We have assumed dissociation profiles for H$_2^{16}$O and $^{28}$SiO of the type
\begin{equation}
\label{equ:diss_SiO}
f(r) = f^\circ \,e^{-(r/r_{\rm e})^p},
\end{equation}
where the standard value of the exponent $p$ is 2 and $r_e$ is the $e$-folding radius, the radius at which the abundance has decreased by a factor $e$
from its initial value.
H$_2^{16}$O is expected to dissociate closer to the star than $^{12}$CO \citep[e.g.][]{Netzer1987}.
\cite{Groenewegen1994} argues that H$_2^{16}$O emission should come from within roughly 100 $R_{\star}$, from considering studies
of OH density profiles in AGB stars. This limit corresponds to an abundance profile with $e$-folding radius of 65 $R_{\star}$, or $1.8 \times 10^{15}$ cm.
For $^{28}$SiO, there are no theoretical estimates that give its dissociation radius in terms of envelope parameters.
That said, \cite{GonzalezDelgado2003}
modelled low-excitation $^{28}$SiO transitions and determined the $e$-folding radius to be $r_{e} = 2.4 \times 10^{15}$ cm for W Hya, which
corresponds to roughly 85 $R_\star$ in the context of our model. At about 200 $R_{\star}$ all of the
$^{28}$SiO will then have disappeared. We adopt these values.  They show that the $^{28}$SiO envelope is
comparable in size to the H$_{2}$O envelope and that both are considerably smaller than the
CO envelope.
Because the $^{28}$SiO transitions modelled by
us are formed mostly deep inside the $^{28}$SiO envelope, we expect the problem of constraining the $^{28}$SiO 
dissociation radius to have little impact on our results.

  \begin{figure}
   \centering
   \includegraphics[width=8.5cm]{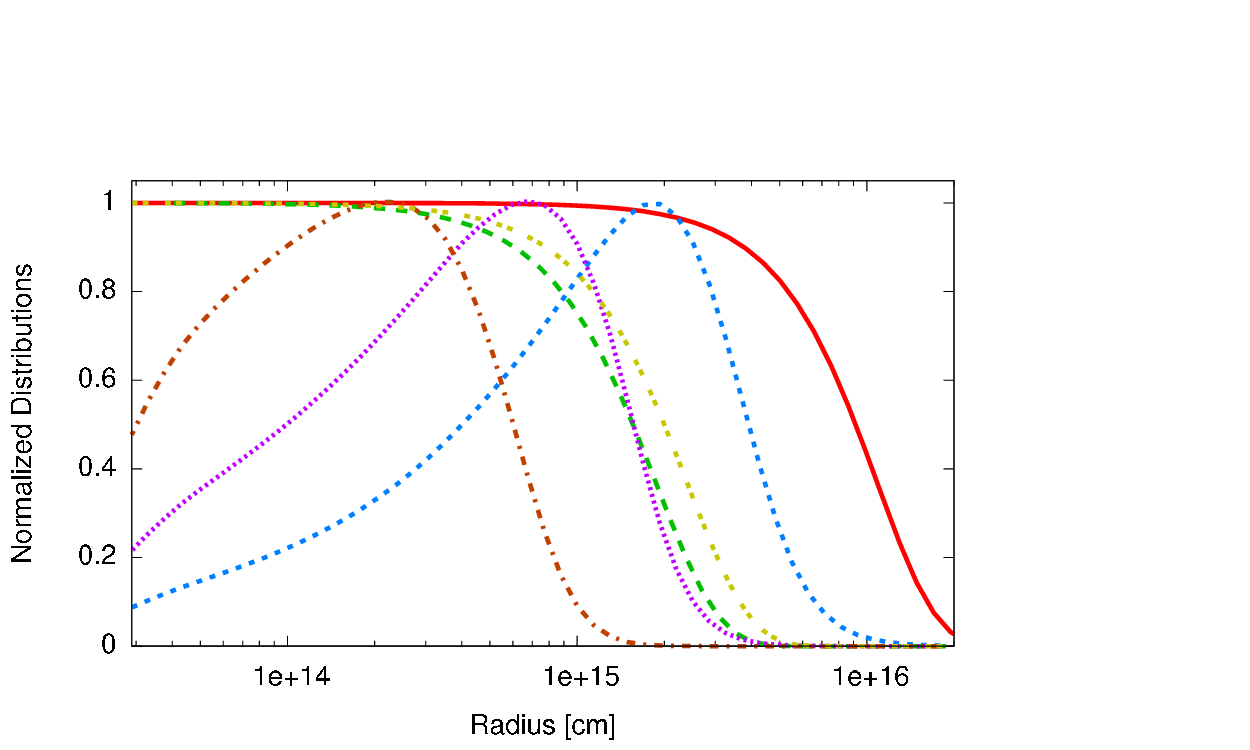}
   \caption{The normalized abundance profile of H$_2^{16}$O (long-dashed green), $^{28}$SiO (double-short-dashed yellow) 
                 and $^{12}$CO (solid red) are compared to the normalized populations of levels $J$ = 16 (dashed-dotted brown), 
                 $J$ = 10 (dotted purple) and $J$ = 6 (short-dashed blue) of $^{12}$CO.}
                \label{fig:regions_probed}
    \end{figure}

In Fig. \ref{fig:regions_probed}, we plot the normalized abundance profiles of $^{12}$CO, $^{28}$SiO and H$_2^{16}$O compared to the 
excitation region of the $^{12}$CO transitions observed by HIFI.  This shows that the H$_2^{16}$O and $^{28}$SiO emissions probe
a relatively small region of the envelope compared to $^{12}$CO emission.

  \begin{figure}   
      \centering
      \includegraphics[width=8.5cm]{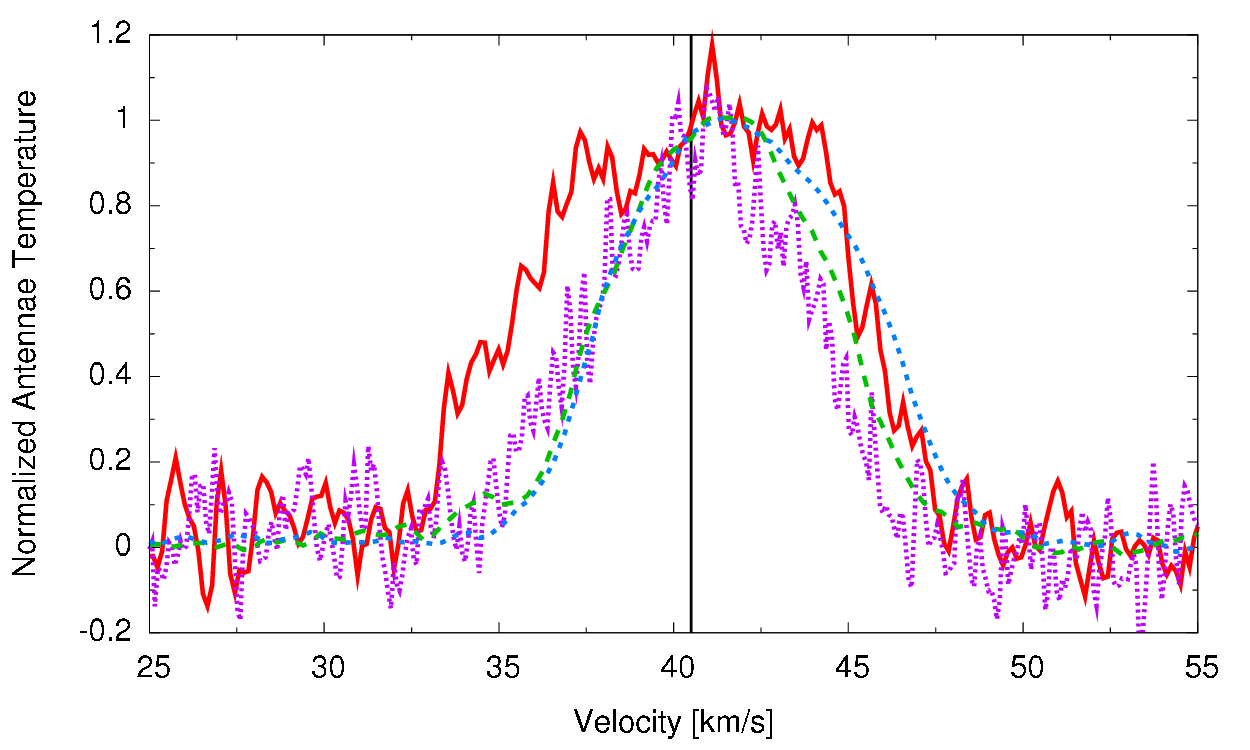}
         \caption{Observed line shapes of the transitions $J$ = 16\,--\,15 (dotted purple) and $J$ =10\,--\,9 
                      (solid red) of $^{12}$CO, $J$ = 14\,--\,13 of $^{28}$SiO (long-dashed green), and 1$_{1,0}$\,--\,0$_{0,1}$
                      of ortho-H$_2^{16}$O (short-dashed blue). The vertical line marks the adopted $\upsilon_{\rm LSR} =
                      40.4$\,km\,s$^{-1}$.}
         \label{fig:comp_shapes}
   \end{figure}

\subsection{Observed line shapes: blue wing absorption}
\label{sec:observed_shapes}

  \begin{figure}[h!]
   \centering
   \includegraphics[width=8.5cm]{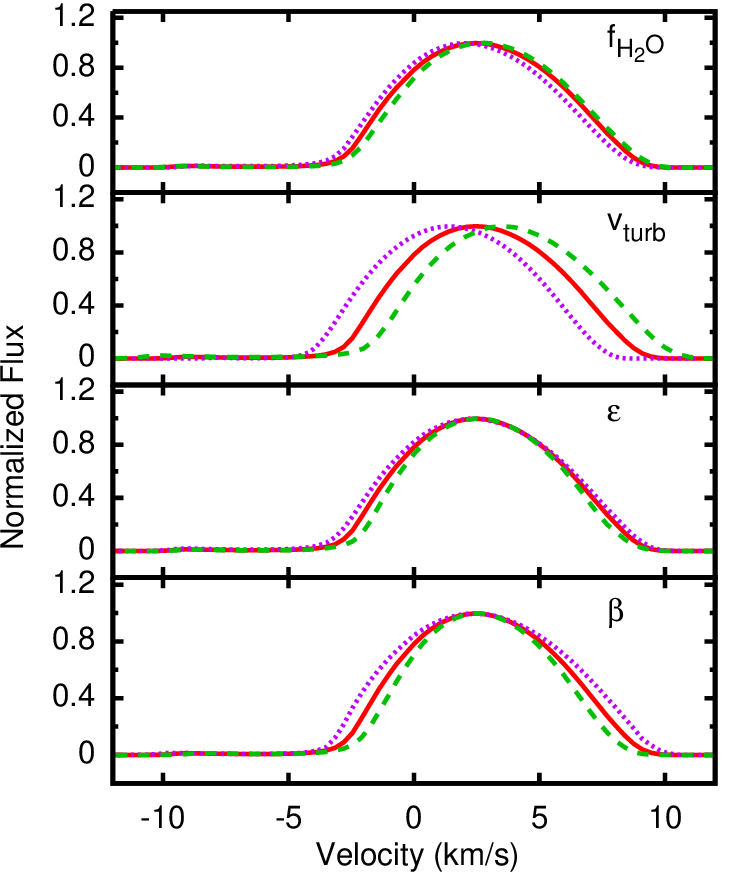}
   \caption{The effects of varying the H$_2^{16}$O abundance, the turbulence velocity, the temperature power law exponent 
                 and the velocity power law exponent on the normalized line strength are shown for o-H$_2^{16}$O ($1_{1,0}-1_{0,1}$).  
                 The standard model is represented by the red line in each panel, it has its values given in Table~\ref{tab:grid_par} and $f_{\rm H_2O}=4\times10^{-4}$.
                 The short-dashed purple and the long-dashed green represent models with respectively: $f_{\rm H_2O}=1\times10^{-4}$ and $1.6\times10^{-3}$ in the first panel;
                 $\upsilon_{\rm turb}=0.8$ and 2.0 km\,s$^{-1}$ in the second panel; $\epsilon=0.2$ and 1.1 in the third panel; and $\beta = 1.5$ and 10 in the fourth panel.
                 All other parameters have the same value as the standard model. We note that varying $\upsilon_{\rm turb}$ has the effect of a line shift.}
\label{fig:lineShift_Par}
   \end{figure}

Fig. \ref{fig:comp_shapes} shows the spectrally resolved profiles of the lowest
excitation transitions of H$_2^{16}$O and $^{28}$SiO, as observed by HIFI, and of the $J$ = 16\,--\,15 and 10\,--\,9
transitions of $^{12}$CO. The adopted local-standard-of-rest velocity of 40.4 km s$^{-1}$ was derived by \cite{Khouri2014} by modelling
the $^{12}$CO and $^{13}$CO transitions. The plotted H$_{2}^{16}$O and $^{28}$SiO lines are expected to form in about
the same region of the envelope as the pure rotational $J$ = 10\,--\,9 line of $^{12}$CO, as they are low-excitation transitions
expected to be formed in the outer parts of the $^{28}$SiO and H$_2^{16}$O envelopes. The $^{12}$CO $J$ = 10\,--\,9 transition 
shows emission up to the same expansion velocity in the red- and blue-shifted wings. The H$_2^{16}$O and $^{28}$SiO transitions, however, show very asymmetrical line profiles
with emission extending to larger velocities in the red-shifted wing than on the blue-shifted wing.
To show the effect of differences in line forming region
(which cannot be the cause of this behaviour, as the lines are selected to form in about the same part of the flow), we
also show the $^{12}$CO $J$ = 16\,--\,15 line profile. This line forms more to the base of the outflow.  The profile is narrower on both sides of line centre.

\cite{Huggins1986} studied the line shapes of transitions $J=2-1$ and $J=1-0$ of $^{12}$CO in IRC\,+10216 in comparison to the line shape of the
optically thinner $^{13}$CO $J=2-1$ line. The authors observed asymmetries between the $^{12}$CO and $^{13}$CO lines and concluded that these
were due to the higher optical depths in the $^{12}$CO lines. They found the effect to be strongly dependent on the turbulence velocity in the line
formation region.
In Fig. \ref{fig:comp_shapes}, the differences between H$_{2}$O and $^{28}$SiO on the one hand and $^{12}$CO $J$ = 10\,--\,9 on the other
hand also arise as a result of differences in line optical depth, and the presence of a turbulence velocity field.  
The $^{12}$CO radial optical depth at line centre is about unity for the transitions discussed.
The Einstein coefficient for spontaneous emission of the H$_2^{16}$O and $^{28}$SiO transitions modelled here are typically
at least an order of magnitude larger than that of $^{12}$CO. The $^{28}$SiO $J$ = 16\,--\,15 and 14\,--\,13 lines 
have even two orders of magnitude higher values.
As a result, higher optical depths are built-up for $^{28}$SiO and H$_2^{16}$O transitions than for $^{12}$CO transitions, even when the different abundances are taken into account.
The effect is important even in the wind of a low mass-loss rate AGB star
such as W Hya, and it reveals itself by the shift in the observed line centre velocity with respect to 
the $\upsilon_{\rm LSR}$ of the source.

In Fig.~\ref{fig:lineShift_Par}, we show that $\upsilon_{\rm turb}$ indeed affects strongly the observed behaviour of H$_2^{16}$O and $^{28}$SiO. 
The o-H$_2^{16}$O ($1_{1,0}-1_{0,1}$) transition is shown for models based on the best $^{12}$CO model obtained in Paper I. 
We varied either the H$_2^{16}$O abundance, the turbulence velocity in the envelope, $\upsilon_{\rm turb}$, the exponent of 
the temperature power law, $\epsilon$, or the exponent of the velocity power law, $\beta$.
Although all these parameters have an impact on the H$_2^{16}$O line shapes or central peak position, it is the turbulence 
velocity that causes a systematic shift of the line centre position. 
We stress that varying these parameters does have a significant effect on
the line strength. To bring out the effects on profile shape and profile shift, we have normalized the lines.
As pointed out by \cite{Huggins1986}, the magnitude of the shift is mainly set by and can be used to determine the value of the turbulence velocity in the line formation region.

\section{H$_{\bf 2}^{\bf 16}$O model}

\label{sec:WHya_H2O}

We calculated models using the parameters found in Paper I for different values of
the ortho- and para-H$_2^{16}$O abundance and the turbulence velocity in the wind. The turbulence velocity was 
included as a free parameter since it was clear from the first calculated H$_2^{16}$O profiles
that the originally assumed value ($\upsilon_{\rm turb}$ = 1.4 km\,s$^{-1}$; see Table~\ref{tab:grid_par}) was too 
large to match the line-centre shifts seen in the line profiles obtained with HIFI (see Sect.~\ref{sec:observed_shapes}).  
The H$_2^{16}$O abundance at a radius $r$ is given by Equation \ref{equ:diss_SiO}. The value adopted by us for the H$_{2}$O dissociation 
radius is that given by \cite{Groenewegen1994}. Our model calculations, however, show that the 
derived H$_2^{16}$O abundance is not strongly affected by the assumed dissociation radius.

\begin{figure*}
   \centering
   \includegraphics{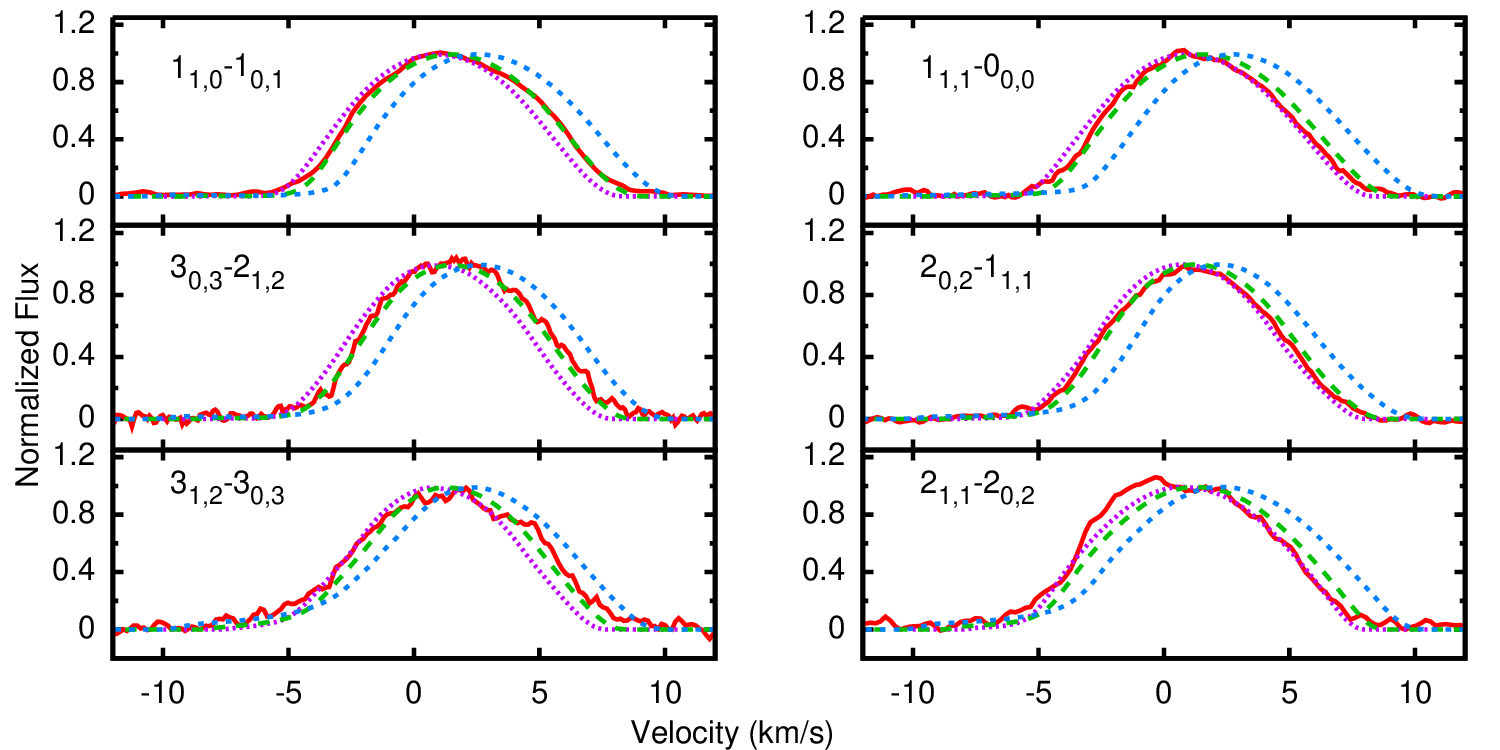}
\caption{Normalized profiles of ortho- and para-H$_2^{16}$O transitions observed by HIFI compared to models with different 
              values of the turbulence velocity and all other parameters kept fixed.
              The solid red line represents the data, and models with turbulence velocities 
              of 0.5, 0.8 and 1.4 km\,s$^{-1}$ are represented by dotted purple, long-dashed green and 
              short-dashed blue lines, respectively. The adopted value for the $\upsilon_{\rm LSR}$ of 40.4 km\,s$^{-1}$$^{-1}$ was subtracted from the observed lines.}
\label{fig:lineShift_water}
   \end{figure*}

The turbulence velocity and the abundance of each spin isomer constitute a degeneracy. 
We can, however, determine with a good accuracy the value of the turbulence velocity from matching the 
line-centre shifts seen by HIFI (see Sect.~\ref{sec:observed_shapes}).  The parameters used in the calculation 
of the H$_2^{16}$O models are listed in Table \ref{tab:waterPar}.  The line profiles computed for different values
of the turbulence velocity are compared to the line shapes observed 
by HIFI in Fig.~\ref{fig:lineShift_water}. The values of the turbulence velocity that best match
the shift seen for ortho-H$_2^{16}$O and para-H$_{2}$O are 0.8 and 0.6 km\,s$^{-1}$, respectively. 
Although the difference in the derived turbulence velocity is small ($\sim$0.2\,km\,s$^{-1}$), the high-quality HIFI data
seems to suggest that the main line-formation region for the o-H$_2$O lines has a slightly different
turbulence velocity than for p-H$_2$O. However, as an uncertainty of 0.1\,km\,s$^{-1}$ in the turbulence velocity, propagates to
an uncertainty of only 5\% at most for the predicted line profiles, we adopt a value of 0.7\,km\,s$^{-1}$ for both spin isomers.

\begin{table}
\caption{Turbulence velocities and abundances relative to H$_{2}$ considered in the models studying the effects of H$_2^{16}$O.}
\label{tab:waterPar}
\centering
\begin{tabular}{c | c}
Parameter & Values \\
\hline\\[-8pt]
$\upsilon_{\rm turb}$ [km\,s$^{-1}$] & 0.5, 0.8, 1.1, 1.4\\
\\
$f$(o-H$_2^{16}$O) & (1, 2, 4, 5, 6, 7, 8, 10, 12, 16) $\times 10^{-4}$ \\ 
\\
$f$(p-H$_2^{16}$O) & (1, 2, 3, 4, 5, 6, 8) $\times 10^{-4}$ \\
\\
\end{tabular}
\end{table}

\begin{figure}
   \centering
   \includegraphics[width=8.5cm]{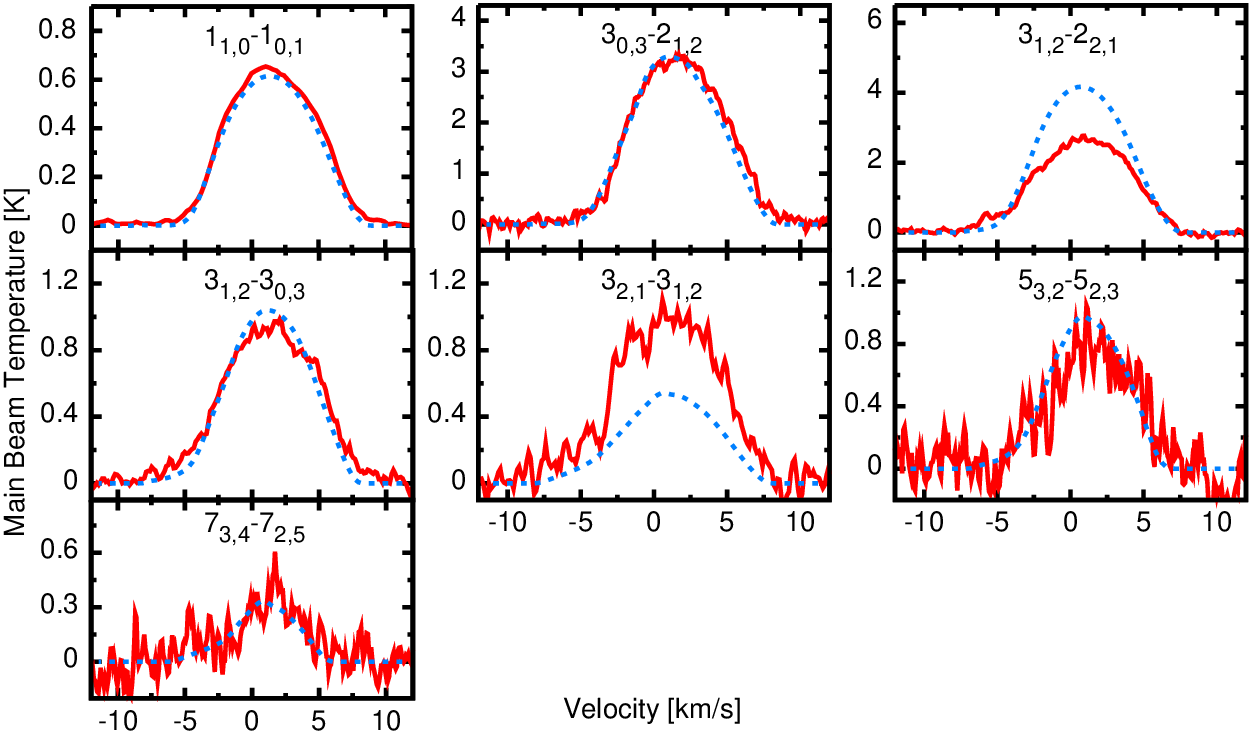}
\caption{Best fit model (dashed-blue line) to the ortho-H$_2^{16}$O lines compared to the lines observed by HIFI (solid-red line). The model has an ortho-H$_2^{16}$O abundance of
$6 \times 10^{-4}$ and a turbulence velocity of 0.7 km\,s$^{-1}$. The adopted value for the $\upsilon_{\rm LSR}$ of 40.4 km\,s$^{-1}$ was subtracted from the observed lines.}
\label{fig:o-water_HIFI}
   \end{figure}

\begin{figure}
   \centering
   \includegraphics[width=8.5cm]{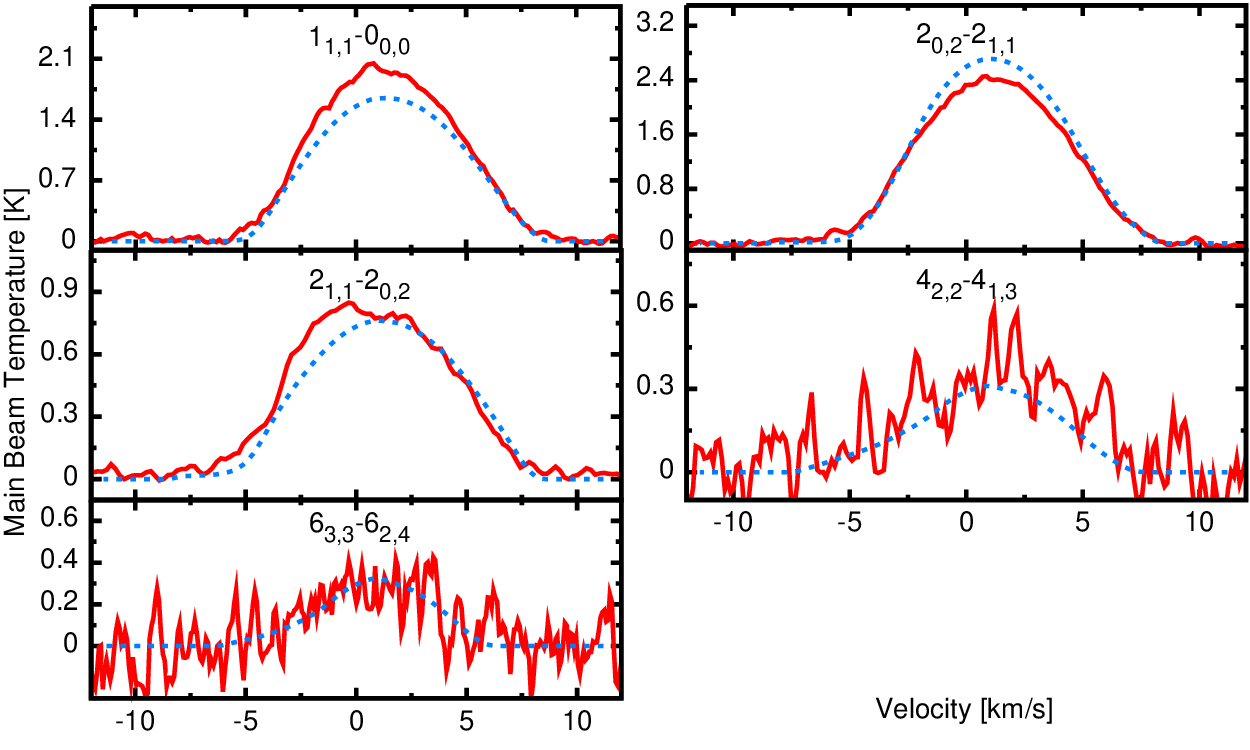}
\caption{Best fit model (dashed-blue line) to the para-H$_2^{16}$O lines compared to the lines observed by HIFI (solid-red line). The model has a para-H$_2^{16}$O abundance of
$3 \times 10^{-4}$ and a turbulence velocity of 0.7 km\,s$^{-1}$. The adopted value for the $\upsilon_{\rm LSR}$ of 40.4 km\,s$^{-1}$ was subtracted from the observed lines.}
\label{fig:p-water_HIFI}
   \end{figure}

After fixing the turbulence velocity, we calculated the reduced-$\chi ^2$ of the fits to 
the PACS and HIFI line fluxes for models with different values of the ortho-H$_2^{16}$O and para-H$_2^{16}$O abundances.
For ortho-H$_2^{16}$O, we get the best result for an abundance of $6 \times 10^{-4}$ for both HIFI and PACS data. 
For para-H$_2^{16}$O, a value of $3 \times 10^{-4}$ fits better the PACS data, while a value between $4 \times 10^{-4}$ 
and $3 \times 10^{-4}$ is the best match for the HIFI observations. Taking the whole dataset into account, the best 
fits for ortho- and para-H$_2^{16}$O are achieved with abundances of $(6^{+3}_{-2} ) \times 10^{-4}$ and $(3^{+2}_{-1}) \times 10^{-4}$, 
respectively. These models are compared to the lines observed by HIFI in Figs.~\ref{fig:o-water_HIFI} and
\ref{fig:p-water_HIFI} for ortho- and para-H$_2^{16}$O, respectively.

\subsection{H$_2$O Isotopologues}

We calculated models for each of the four isotopologues of H$_2$O, ortho-H$_2^{17}$O and H$_2^{18}$O and 
para-H$_2^{17}$O and H$_2^{18}$O, considering both a turbulence velocity of 0.7\,km\,s$^{-1}$
and different values for the abundance of each isotopologue.  The optical depths of the transitions are
considerably smaller than the corresponding ones of the main isotopologues, causing the derived abundances
to be roughly independent of $\upsilon_{\rm turb}$. The uncertainty on $\upsilon_{\rm turb}$ propagates to an uncertainty
of a few percent in the line fluxes of the rarer isotopologues.

\begin{table}
\caption{Best fit results for the abundances of the isotopologues of H$_2$O. The super- and subscript give the 1-$\sigma$ uncertainties in the determined values.}
\label{tab:waterIsotBest}
\centering
{
\renewcommand{\arraystretch}{1.5}
\begin{tabular}{c | r }
Molecule & Best \\
\hline\\[-8pt]
para-H$_2^{17}$O & $2^{+1}_{-1} \times 10^{-7}$ \\ 
para-H$_2^{18}$O & $1.6^{+0.4}_{-0.6} \times 10^{-6}$\\ 
ortho-H$_2^{17}$O & $6^{+4}_{-2} \times 10^{-7}$\\ 
ortho-H$_2^{18}$O & $3^{+1}_{-1} \times 10^{-5}$\\ 
\end{tabular}
}
\end{table}

\begin{table}
\caption{Derived isotopologic ratios. The super and subscript give the 1-$\sigma$ uncertainties in the determined values.}
\label{tab:waterRatios}
\centering
{
\renewcommand{\arraystretch}{1.5}
\begin{tabular}{c c }
Molecules & Ratio$^{\rm ~upper\,limit}_{\rm ~lower\,limit}$\\
\hline
o-H$_2^{16}$O/o-H$_2^{17}$O & 1000\,$^{+1200}_{-600}$ \\ 
(o-H$_2^{16}$O/o-H$_2^{18}$O) & 20\,$^{+25}_{-8}$ \\ 
p-H$_2^{16}$O/p-H$_2^{17}$O & 1500\,$^{+2500}_{-800}$ \\ 
p-H$_2^{16}$O/p-H$_2^{18}$O & 190\,$^{+210}_{-90}$ \\ 
o-H$_2^{16}$O/p-H$_2^{16}$O & 2\,$^{+2.5}_{-0.8}$ \\ 
o-H$_2^{17}$O/p-H$_2^{17}$O & 3\,$^{+7}_{-1.5}$ \\ 
(o-H$_2^{18}$O/p-H$_2^{18}$O) & 19\,$^{+21}_{-9}$ \\ 
p-H$_2^{18}$O/p-H$_2^{17}$O & 8\,$^{+12}_{-5}$ \\ 
(o-H$_2^{18}$O/o-H$_2^{17}$O) & 50\,$^{+50}_{-30}$ \\ 
\hline

\end{tabular}
}
\tablefoot{The ratios between parentheses were not considered when deriving the final values. }
\end{table}

  \begin{figure}
   \centering
   \includegraphics[width=8.5cm]{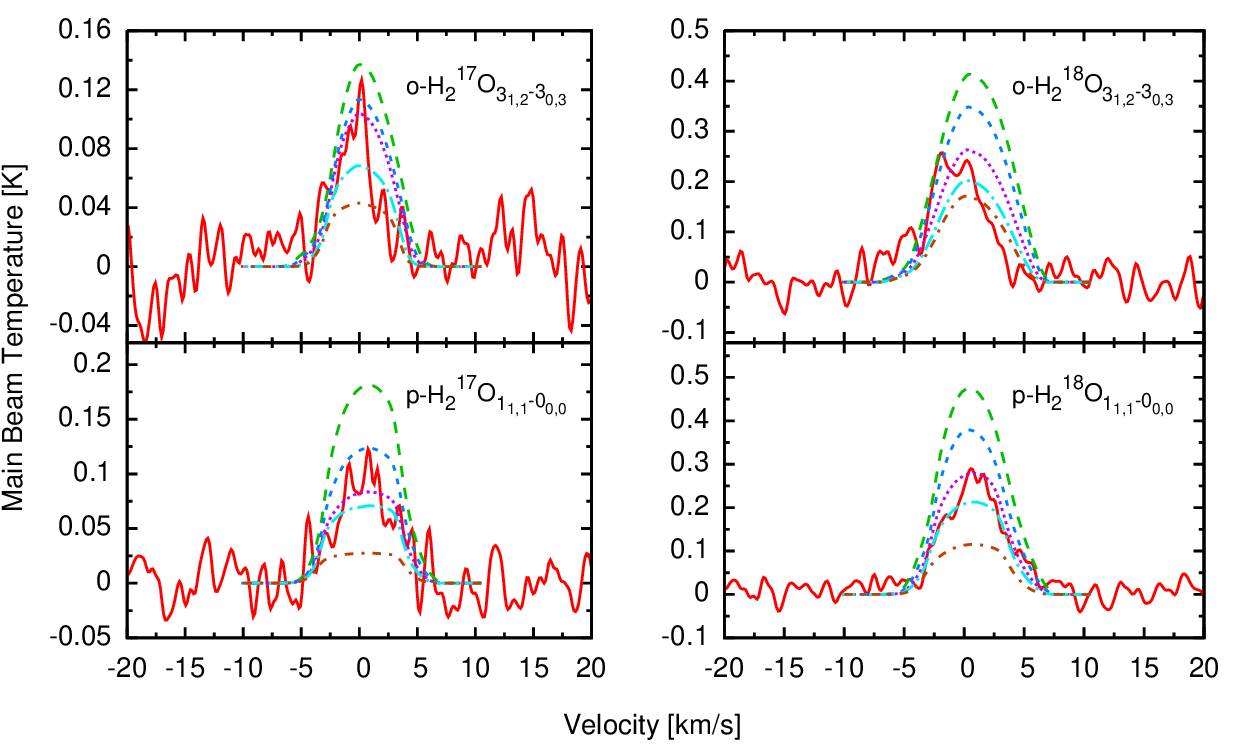}
\caption{The solid red line represents the HIFI data, the dashed green, short-dashed blue, dotted purple, 
              dashed-dotted light-blue and short-dashed-dotted brown lines represent, respectively, models with 
              abundances relatively to H$_2$ of: (4, 2, 1.2, 0.4, 0.2) $\times 10^{-6}$ for 
              for ortho-H$_2^{17}$O;  (1.2, 0.8, 0.4, 0.2, 0.12) $\times 10^{-4}$ 
               for ortho-H$_2^{18}$O;  (8, 4, 2, 1.2, 0.4) $\times 10^{-7}$ 
               for para-H$_2^{17}$O; and (8, 4, 2, 1.2, 0.4) $\times 10^{6}$ 
               for para-H$_2^{18}$O. We adopted a value of 0.7 km\,s$^{-1}$ for the turbulence velocity in these calculations.
               The adopted value for the $\upsilon_{\rm LSR}$ of 40.4 km\,s$^{-1}$ was subtracted from the observed lines.}
\label{fig:isot_water}
   \end{figure}
   
The models are compared to the observations in Fig.~\ref{fig:isot_water} and the best abundance values of 
each isotopologue relative to H$_2$ are given in Table \ref{tab:waterIsotBest}. The best models were selected 
by comparing the integrated line fluxes with the observations. Then, we computed the different 
isotopologic ratios given in Table \ref{tab:waterRatios}, which allow us to derive the $^{16}$O/$^{17}$O, the $^{16}$O/$^{18}$O 
isotopic ratios and the ortho-to-para ratio. We note, however, that the
line shapes of the two ortho-H$_2$O isotopologues are not well reproduced, particularly that of o-H$_2^{18}$O.
The single observed line of o-H$_2^{18}$O is shifted with respect to the modelled transitions and very high values of
the o-H$_2$$^{18}$O abundance are needed to predict the observed line strength.
Therefore, the values derived for the $^{16}$O/$^{18}$O ratio and for the H$_2$O ortho-to-para ratio based on this line do not
agree with those derived using any other pair of observed lines. Because of this mismatch, we have not included the values derived from the
o-H$_2$$^{18}$O line in the ortho-to-para and isotopic ratios calculations. Modifying the dissociation radius and turbulence velocity
does not improve the fitting of this transition.
Discarding the o-H$_2$$^{18}$O line, we obtain: ortho-to-para = 2.5\,$^{+2.5}_{-1.0}$, $^{16}$O/$^{17}$O = 1250\,$^{+750}_{-450}$ and
$^{16}$O/$^{18}$O = 190\,$^{+210}_{-90}$ (see also Table~\ref{tab:waterRatios}).

\subsection{Consequences of a lower $\upsilon_{\rm turb}$ for the $^{12}$CO lines}

\begin{figure}
   \centering
   \includegraphics[width=8.5cm]{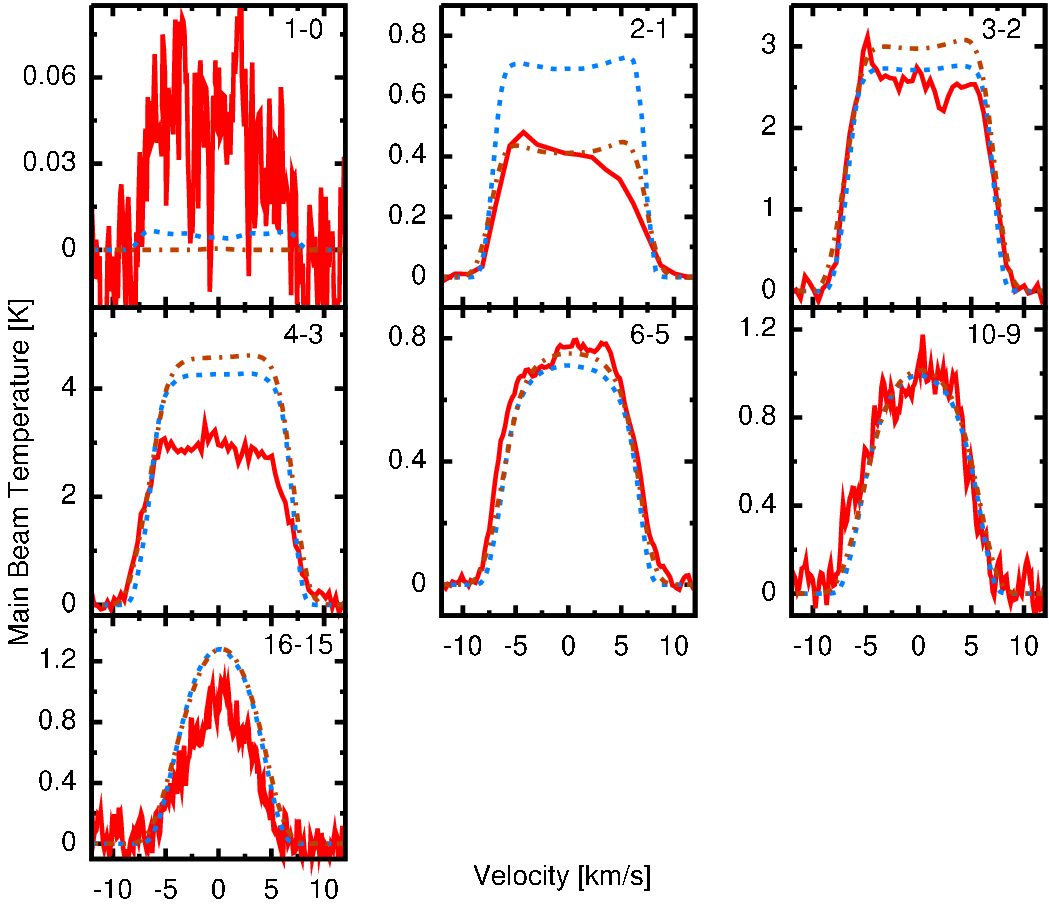}
   \caption{A model for the $^{12}$CO lines with $\upsilon_{\rm turb} = 0.7$ km\,s$^{-1}$ (short-dashed blue line) is compared to the best model 
                found in Paper I ($\upsilon_{\rm turb} = 1.4$ km\,s$^{-1}$; dotted-dashed brown line). The observed $^{12}$CO lines are shown
                in red. The adopted value for the $\upsilon_{\rm LSR}$ of 40.4 km\,s$^{-1}$ was subtracted from the observed lines. The lines
                $J=16-15$, $10-9$ and $6-5$ were observed with HIFI, $J=4-3$ and $3-2$ with APEX, $J=2-1$ with SMT and $J=1-0$ with SEST.}
   \label{fig:CO_lowerStochV}
   \end{figure}

The turbulence velocity that resulted from the $^{12}$CO analysis in Paper~I was 1.4 km\,s$^{-1}$. However, this value was
not strongly constrained.
Although this higher value is better at reproducing the $^{12}$CO line wings, the differences between models for 0.6--0.8 and
1.4 km\,s$^{-1}$ are small. In Fig.~\ref{fig:CO_lowerStochV} we show a comparison between the $^{12}$CO model from 
Paper I and a model with $\upsilon_{\rm turb} = 0.7$ km\,s$^{-1}$.
The change in the value of the turbulence velocity only has an impact on the total line flux of the transitions having
 $J_{\rm up} <6$, most notably that of $J$ = 2\,--\,1.
When a lower value of the turbulence velocity is used, the model predictions for the $J$ = 1\,--\,0 and 2\,--\,1 transitions get somewhat 
stronger, while those for the 3\,--\,2 to 6\,--\,5 transitions get somewhat weaker. The emission from line $J$ = 2\,--\,1 
could be decreased by considering an even smaller $^{12}$CO dissociation radius than the one obtained in Paper I,
but this will not resolve the discrepancy in the $J$ = 1\,--\,0 line.  The poor fit to the second line is the
main shortcoming of our $^{12}$CO model.
We conclude that a lower value for the turbulence velocity
does not affect significantly the quality of our fit, except for the very low-excitation
lines $J=2-1$ and $J=1-0$, the second of which was also poorly reproduced by our original $^{12}$CO model.
Furthermore, we can expect the turbulence velocity to be different in the formation regions of
H$_2^{16}$O and $^{28}$SiO and of the low-excitation $^{12}$CO lines, since the $^{12}$CO envelope is significantly larger
than that of H$_2^{16}$O and $^{28}$SiO. However, since it is not possible to determine a precise value for the
turbulence velocity from the $^{12}$CO lines, we cannot draw any conclusion on changes in this parameter
between the H$_2^{16}$O and $^{28}$SiO envelopes and the $^{12}$CO outer envelope.

\subsection{Reproducing the PACS spectrum}

\begin{figure*}
   \centering
   \includegraphics[width=18cm]{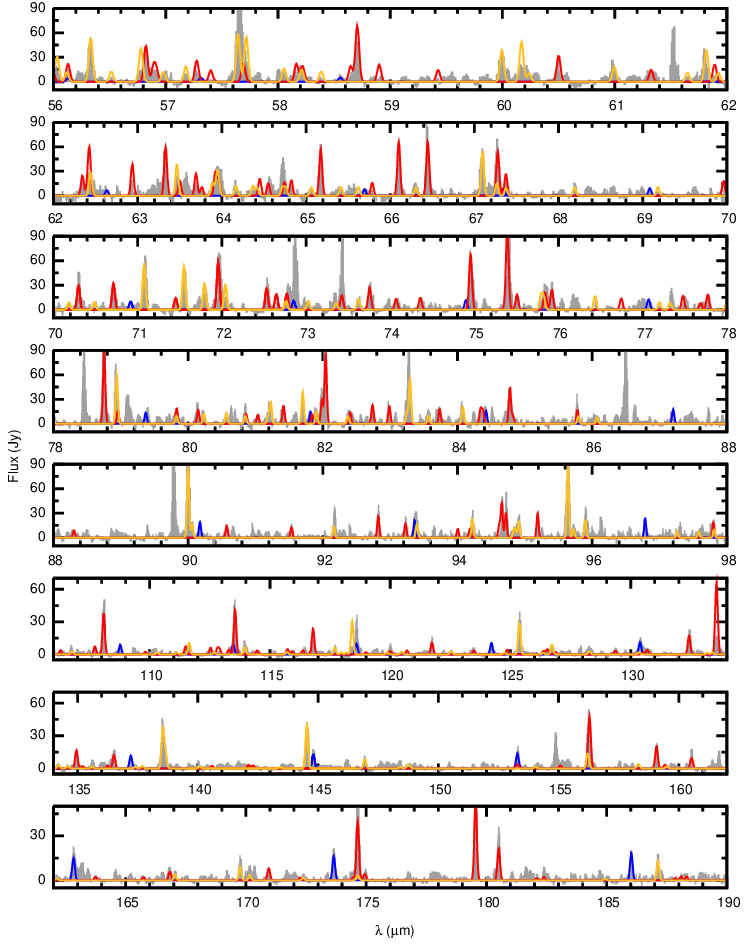}
\caption{The PACS spectrum (represented by the grey-filled histogram) is compared to our best model, with the parameters given in Table \ref{tab:grid_par}
but with $\upsilon_{\rm turb}$ of 0.7 km\,s$^{-1}$ and with ortho- and para-H$_2^{16}$O abundances of
$6\times10^{-4}$ and $3\times10^{-4}$, respectively. The $^{12}$CO model is shown by the full-blue line, the ortho-H$_2^{16}$O, by the full-red line and the para-H$_2^{16}$O,
by the full-yellow line.}
\label{fig:PACS_full}
   \end{figure*}

In Fig. \ref{fig:PACS_full}, we compare our $^{12}$CO and ortho- and para-H$_2^{16}$O models to the PACS spectrum of W\,Hya.
The $^{28}$SiO lines modelled do not contribute significantly to the PACS spectrum and were not included in the plot.
The vast majority of the prominent lines seen in the spectrum can be accounted for by our H$_2^{16}$O model. A small fraction of
the strong lines, however, is not predicted. These lines have peaks around 61.52, 72.84, 78.47, 79.12,
86.52, 89.78, 154.88 and 163.12 $\mu$m. The lines observed at 79.12 and 163.12 $\mu$m can be associated with OH transitions \citep{Sylvester1997}.
Those at 78.47, 86.52 and 89.78 $\mu$m might be explained by $^{28}$SiO maser emission (Decin et al., \textit{in prep.}). We were not able to identify the lines observed
at 61.52, 72.84 and 154.88 $\mu$m.

\section{Model for gas-phase $^{\bf 28}$SiO emission}
\label{sec:SiO}

In Paper I, we assumed a standard value for the $^{12}$CO abundance relative to H$_2$ of $4 \times 10^{-4}$. 
Were all carbon and silicon used to form $^{12}$CO and $^{28}$SiO, respectively, and solar composition assumed, then the $^{12}$CO-to-$^{28}$SiO abundance ratio would be roughly 8.3,
corresponding to an $^{28}$SiO photospheric abundance of $4.8\times10^{-5}$.

In order to establish the location in the outflow where silicate dust particles condense, and the fraction of 
the $^{28}$SiO that is converted from the molecular phase to the solid phase in the inner wind, we modelled the $^{28}$SiO line emission detected by
{\em Herschel}.
The broad spectral coverage of PACS and SPIRE provides a series of lines with upper-level energies ranging
from 137 to 1462 K above the ground state,
covering the region where silicates are expected to condense \citep{Gail1999} and, consequently, where a decrease in the $^{28}$SiO abundance should be seen.

We compare these data to a grid of models in which the photospheric $^{28}$SiO abundance relative to H$_{2}$,
$f_{\rm SiO}^{\circ}$, the condensation fraction of $^{28}$SiO in solid material, $f_{\rm cond}$, and radius at which this
happens, $R_{\rm cond}$, are varied.  For $f_{\rm SiO}^{\circ}$ we apply (10, 8, 6, 5, 4, 2) $\times 10^{-5}$.
For $f_{\rm cond}$ we adopt 0, 0.35, 0.65, and 0.90.  The first value implies no condensation of solids.  For 
each of the abundance profiles in which condensation was considered, we take
$R_{\rm cond}$ to be either 5, 10, or 20 $R_{\star}$. This adds up to a total of sixty models.
We note that models without condensation are equivalent to models with a high $^{28}$SiO abundance in which condensation
occurs deep in the envelope, at about 2\,$R_{\star}$, as our models are not sensitive to the abundance at radii smaller than about 5\,$R_{\star}$.
To give an example, 
a model with an initial abundance of $4 \times 10^{-5}$ is analogous to a model with $f_{\rm SiO}^{\circ}
= 6 \times 10^{-5}$ and $f_{\rm cond} = 0.33$ in which condensation occurs close to the star.

\begin{figure}
   \centering
   \includegraphics[width=8.5cm]{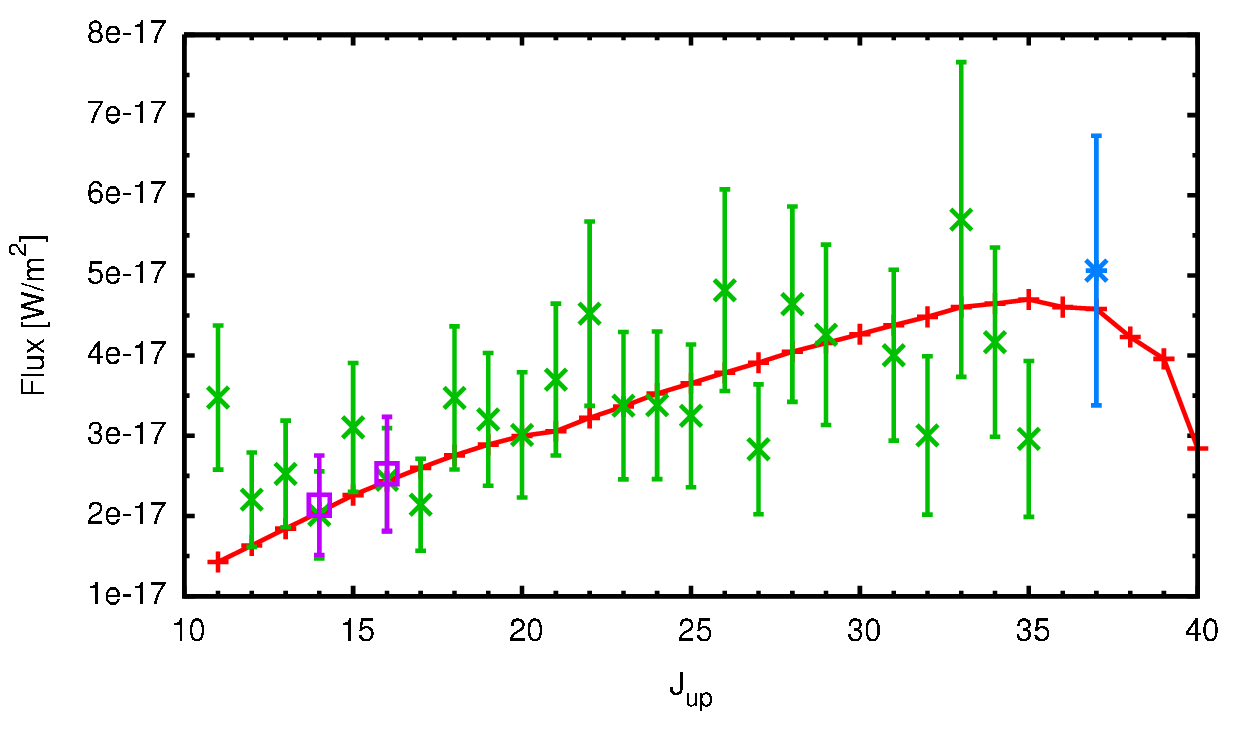}
\caption{The best model for the $^{28}$SiO line emission (red line and crosses), with $f^{\rm SiO}_\circ$ = $4\times10^{-5}$ and $f_{\rm cond}=0$,
is compared to the line fluxes observed by SPIRE (green), PACS (blue) and HIFI (purple).}
\label{fig:SiO_LineFluxes}
   \end{figure}

As shown in Table \ref{tab:sio_chi}, the models were ranked based on the calculated reduced-$\chi^2$ fit to the line fluxes obtained 
with SPIRE, HIFI and PACS, listed in Table~\ref{tab:obs_SiO} of the appendix.  The ten best models have $^{28}$SiO abundances 
between 2.5 $\times 10^{-5}$ and 4.0 $\times 10^{-5}$ relative to H$_2$ in the region between 10 and 100 
$R_\star$, i.e. in the region where the $^{28}$SiO emission originates (see Fig.~\ref{fig:regions_probed}).
The condensation radius of $^{28}$SiO is not strongly constrained. However, condensation at 10 $R_{\star}$ or less 
seems preferred over condensation at 20 $R_{\star}$.
In the top-fifteen-ranked models,
those with a condensation fraction of 0.65 all have higher photospheric $^{28}$SiO abundance than expected on the basis of a 
solar carbon-to-silicon ratio. If the photospheric $^{28}$SiO abundance is assumed to be solar ($f^{\rm SiO}_\circ$ between
4.0 and 6.0$\,\times 10^{-5}$), a condensation fraction of 0.35 or less is preferred. Furthermore, the slope seen in Fig. \ref{fig:SiO_LineFluxes}
for the observed line fluxes distribution in terms of $J$ is much better reproduced by models with no or very little condensation.

\begin{figure}
   \centering
   \includegraphics[width=8.5cm]{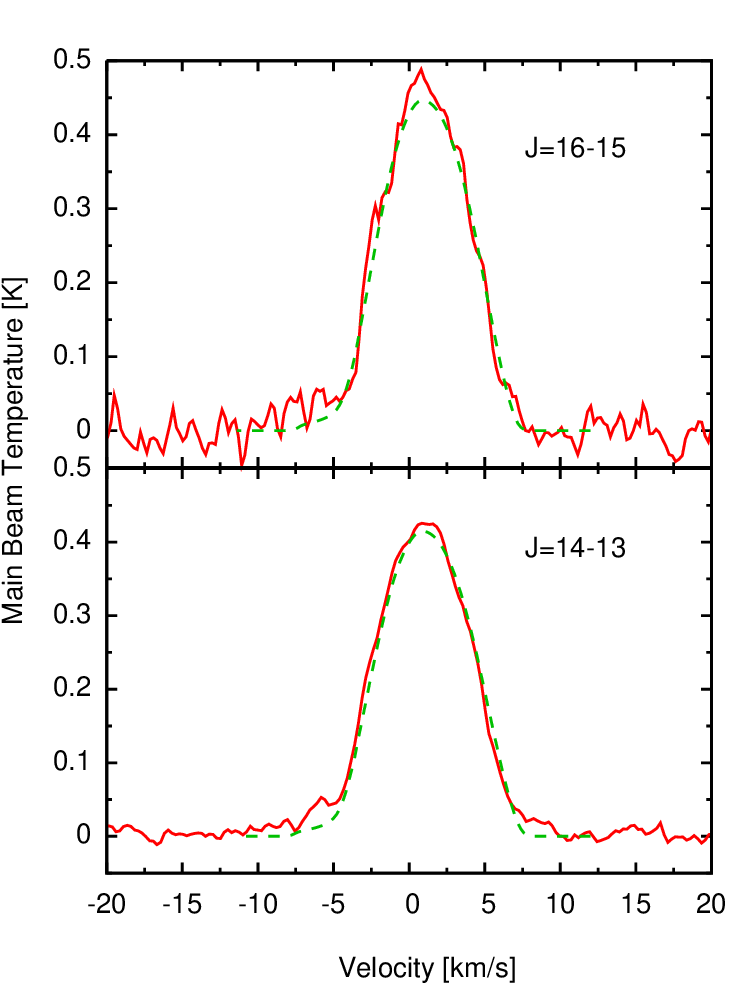}
\caption{The best $^{28}$SiO model (green-dashed line) is compared to the line shapes observed by HIFI (solid red line).
	The adopted value for the $\upsilon_{\rm LSR}$ of 40.4 km\,s$^{-1}$ was subtracted from the observed lines.}
\label{fig:SiO_LineShapes}
   \end{figure}

The fit to the $J$ = 14\,--\,13 and 16\,--\,15 $^{28}$SiO lines observed by HIFI, shown in Fig. \ref{fig:SiO_LineShapes}, is consistent with $\upsilon_{\rm turb} = 0.7\pm0.1$ km\,s$^{-1}$.

\begin{table}[h!]
\caption{Calculated reduced-$\chi^2$ of the fit to the line fluxes of the set of $^{28}$SiO lines listed in 
	      Table~\ref{tab:obs_SiO} of the appendix. 
              The models are listed in order of fit quality, with the best fit at the top.}
\label{tab:sio_chi}
\centering
\begin{tabular}{r c r c}
$f_{\rm SiO}^\circ$/$10^{-5}$ & $f_{\rm cond}$ & R$_{\rm cond}$ [R$_\star$] & red-$\chi^2$ \\
\hline
4.0 & 0 & - & 1.08\\
6.0 & 0.35 & 5 & 1.11\\
10.0 & 0.65 & 5 & 1.18\\
5.0 & 0 & - & 1.20\\
5.0 & 0.35 & 5 & 1.23\\
8.0 & 0.35 & 5 & 1.29\\
6.0 & 0.35 & 10 & 1.30\\
5.0 & 0.35 & 10 & 1.35\\
5.0 & 0.35 & 20 & 1.38\\
4.0 & 0.35 & 20 & 1.40\\
8.0 & 0.65 & 5 & 1.44\\
6.0 & 0 & - & 1.55\\
6.0 & 0.35 & 20 & 1.55\\
4.0 & 0.35 & 5 & 1.57\\
4.0 & 0.35 & 10 & 1.59\\
8.0 & 0.35 & 10 & 1.60\\
10.0 & 0.65 & 10 & 1.72\\
8.0 & 0.65 & 10 & 1.78\\
10.0 & 0.35 & 5 & 1.82\\
5.0 & 0.65 & 20 & 1.84\\
4.0 & 0.65 & 20 & 1.91\\
6.0 & 0.65 & 20 & 1.92\\
6.0 & 0.65 & 5 & 2.03\\
6.0 & 0.65 & 10 & 2.07\\
2.0 & 0 & - & 2.19\\
10.0 & 0.35 & 10 & 2.21\\
8.0 & 0.35 & 20 & 2.25\\
5.0 & 0.65 & 10 & 2.35\\
8.0 & 0.65 & 20 & 2.39\\
5.0 & 0.65 & 5 & 2.51\\
5.0 & 0.90 & 20 & 2.58\\
4.0 & 0.90 & 20 & 2.59\\
2.0 & 0.35 & 20 & 2.62\\
8.0 & 0 & - & 2.66\\
6.0 & 0.90 & 20 & 2.70\\
4.0 & 0.65 & 10 & 2.76\\
10.0 & 0.65 & 20 & 3.06\\
2.0 & 0.65 & 20 & 3.08\\
2.0 & 0.35 & 10 & 3.15\\
8.0 & 0.90 & 20 & 3.16\\
4.0 & 0.65 & 5 & 3.17\\
10.0 & 0.35 & 20 & 3.25\\
2.0 & 0.35 & 5 & 3.46\\
2.0 & 0.90 & 20 & 3.51\\
10.0 & 0.90 & 10 & 3.65\\
10.0 & 0.90 & 20 & 3.77\\
8.0 & 0.90 & 10 & 3.93\\
10.0 & 0 & - & 4.15\\
10.0 & 0.90 & 5 & 4.16\\
6.0 & 0.90 & 10 & 4.32\\
2.0 & 0.65 & 10 & 4.43\\
5.0 & 0.90 & 10 & 4.58\\
8.0 & 0.90 & 5 & 4.87\\
4.0 & 0.90 & 10 & 4.93\\
2.0 & 0.65 & 5 & 5.48\\
6.0 & 0.90 & 5 & 5.79\\
2.0 & 0.90 & 10 & 6.15\\
5.0 & 0.90 & 5 & 6.36\\
4.0 & 0.90 & 5 & 7.04\\
2.0 & 0.90 & 5 & 8.95\\
\hline
\end{tabular}
\end{table}

\section{Discussion}
\label{sec:disc}

\subsection{The turbulence velocity}

The turbulence velocity probed by a given line is that of the region where the line is excited and from which 
photons can escape.  The lines observed by HIFI for the ortho- and para-H$_2^{16}$O transitions have similar 
excitation energies, hence they may be expected to form in a similar part of the outflow if the spin isomers themselves
occupy the same region.

We found a small difference in the turbulent velocity value that predicts best the observed line shapes of the
ortho- and para-H$_2$O, 0.8 km\,s$^{-1}$ and 0.6 km\,s$^{-1}$ respectively. The $\upsilon_{\rm turb}$ derived
by \cite{Khouri2014} based on the $^{12}$CO line shapes was $1.4 \pm 1.0$\,km\,s$^{-1}$.
Since the low-excitation CO lines probe the outer parts of the wind, this could be another indication of a turbulent velocity gradient.
However, the uncertainty in the $\upsilon_{\rm turb}$ values from CO are sizable. Moreover, the observed 
line shapes of both H$_2$O spin isomers are not perfectly fitted for
any value of $\upsilon_{\rm turb}$.
Although 
these diagnostics
 suggest a gradient, GASTRoNOoM can only calculate models with a constant turbulent velocity
and therefore we cannot test this possibility at the moment.  We adopt $\upsilon_{\rm turb} =  0.7\pm0.1$ km\,s$^{-1}$.

\cite{Maercker2009} found
the dissociation radius and the shape of the abundance profile to also have an impact on the line shapes. We discuss our assumptions
for these properties in Section \ref{sec:photodis}.
We have tested the impact of decreasing the dissociation radius on the line shapes. The lines become narrower, the red-wing being more strongly affected.
The dissociation radii and the abundance profile have an impact on the line shapes and their peak position and add to the uncertainty on the
determined turbulence velocity. However, if the abundance profile of ortho- and para-H$_2^{16}$O are similar, there should be no relative difference
between the line shapes of the two spin isomers.

\subsection{Isotopic ratios and evolutionary status of W\,Hya}

During the evolution of low- and intermediate-mass stars leading up to the AGB and on the AGB, the abundances 
of the two minor oxygen isotopes are expected to vary considerably, especially due to the first- and third-dredge-up
events.  The effect of the first-dredge-up on the $^{16}$O/$^{17}$O surface isotopic ratio is found 
to depend quite sensitively on the initial stellar mass, the values of this ratio after this first-dredge-up event is a steeply decreasing
function of stellar mass for stars with main sequence mass between 1 and 3 M$_\odot$ \citep[e.g.][]{Boothroyd1994,Lattanzio1997,Palmerini2011,Charbonnel2010}. During the 
third-dredge-up, the surface oxygen isotopic ratios are expected to change only if hot bottom burning is 
active \citep[e.g.][]{Busso1999,Charbonnel2010,Karakas2011}.  In this case, the abundance of $^{18}$O 
drops strongly while that of $^{17}$O remains almost unchanged. \cite{Justtanont2013} recently lend 
support to this prediction by presenting observations that clearly show that the abundance of
H$_2$$^{17}$O is considerably higher than that of H$_2$$^{18}$O for a set of OH/IR stars observed with HIFI.

The isotopic ratios determined for W\,Hya, $^{16}$O/$^{17}$O = 1250\,$^{+750}_{-450}$ and 
$^{16}$O/$^{18}$O = 190\,$^{+210}_{-90}$, are lower than the solar values of ($^{16}$O/$^{17}$O)$_{\odot}$
= 2600 and ($^{16}$O/$^{18}$O)$_{\odot}$ = 500.
The
observed value of 1250 implies that W\,Hya had an initial mass of about 1.5 M$_\odot$. Evolutionary models show the $^{16}$O/$^{17}$O ratio after
the first dredge-up to be independent of metallicity.
Since the $^{16}$O/$^{17}$O surface ratio is such a steep function of initial mass, the initial mass of W\,Hya
would be constrained to be within 1.3 and 1.7 M$_\odot$. Such a star would reach the AGB phase in about 3 gigayears.
If W\,Hya has less metals than the Sun, the determined value for the $^{16}$O/$^{17}$O surface ratio would also
be consistent with it having an initial mass of more than 4 M$_\odot$. It is unlikely, however, that W\,Hya is either
metal-poorer than the Sun or so massive.

The value for the $^{16}$O/$^{18}$O ratio does not agree with what is found with available evolutionary models. The observed 
value of 190 is lower than that of the Sun. All models predict this ratio to be a weak function of mass and to increase during
evolution, therefore the observed value cannot be reconciled with predictions. Interestingly, \cite{Decin2010} determine 
the $^{16}$O/$^{18}$O ratio of the also oxygen-rich AGB star IK Tau to be 200, a value that is very close to 
the value determined by us for W\,Hya.

One solution to this problem may be that both W\,Hya and IK\,Tau are richer in metals than the Sun, since 
the $^{16}$O/$^{18}$O ratio is expected to be inversely proportional to metal content \citep{Timmes1995}.
The observed isotopic ratio then requires that W\,Hya and IK Tau are about twice as rich in metals than is 
the Sun.
If that is not the case, and the low $^{16}$O/$^{18}$O ratio is confirmed for these two objects, our findings
would imply that the evolution of the $^{18}$O surface abundance up to the AGB stage is not yet well understood. 
However, the uncertainty associated with the $^{16}$O/$^{18}$O measurement does not allow one to draw a firm
conclusion on this matter at the moment.

\subsection{$^{\bf 28}$SiO condensation}

If we consider a photospheric $^{28}$SiO abundance expected for solar composition ($f_{\circ}^{\rm SiO} = 4.8\times10^{-5}$), 
a dust mass-loss rate of $2.8 \times 10^{-10} M_{\odot}{\rm yr}^{-1}$ and the silicon-bearing grains to consist of olivine silicates (MgFeSiO$_{4}$) (see Paper I), 
our model requires about one-third of the silicon atoms in the wind of W\,Hya to be in dust grains.
We thus expect to see a decrease in the $^{28}$SiO abundance from $4.8\times10^{-5}$ to $3.2\times10^{-5}$
over the region where silicates condense.
Furthermore, our dust model predicts that silicates are formed in W\,Hya's wind as close as $\sim$5 stellar radii 
(or 10 AU) to the surface.  Observations carried out by \cite{Zhao-Geisler2011} with
MIDI/VLTI set a lower limit on the silicate formation radius at 28 photospheric radii (equivalent to 50 AU or 0.5 arcsec). That corresponds to 22 R$_{\star}$
(equivalent to $\sim 40$ AU)
in our model, when we correct for the different distance adopted by them.
Furthermore, aperture-masked polarimetric interferometry observations
carried out by \cite{Norris2012} reveal a close-in halo of large transparent grains in W\,Hya. The composition of the
grains could not be determined, but if this material also contains silicon it has to be considered in the silicon budget.

Our molecular-line-emission calculations indicate that $^{28}$SiO does not suffer from strong depletion in W\,Hya's wind.
Furthermore, an $^{28}$SiO abundance of $(3.3 \pm 0.8) \times 10^{-5}$ between 10 and 100 R$_\star$
is required in order to reproduce the observed $^{28}$SiO lines. Adding to that
the silicon that is in our dust model, which corresponds to an abundance of $1.6 \times 10^{-5}$, we reach a 
total silicon abundance of $4.9 \times 10^{-5}$.  This is very close to the abundance expected
based on a solar silicon-to-carbon ratio, i.e. $4.8\times 10^{-5}$ for $^{28}$SiO.
Models with condensation occurring at radii equal to, or smaller than, 10 stellar radii are preferred to those with condensation
at 20 stellar radii but we are not able to determine the condensation radius based in our data.
One could expect condensation to happen over a few or even tens of stellar radii, a scenario not explored in our calculations.
Despite the good agreement regarding the silicon budget, we note that our dust model was based on the one obtained by
\cite{Justtanont2005} and that we have not studied the dust envelope in detail, as considering  different dust species and/or distribution.
Furthermore, the present dust model does not agree with
the observations carried out by \cite{Zhao-Geisler2011}, which show that silicates do not condense closer than about 40 AU.
We will analyse in depth the dust envelope of W\,Hya under the light of the gas-phase wind model presented here in an upcoming study. 

Regarding the outer $^{28}$SiO envelope, \cite{GonzalezDelgado2003} modelled the $^{28}$SiO pure rotational emission of the ground-vibrational state and 
obtained an $^{28}$SiO abundance of $1.5\times 10^{-5}$. The authors, however, studied the $^{28}$SiO abundance relative
to $^{12}$CO mainly in a statistical way. They compared their models to low-excitation transitions, 
$J$ = 2\,--\,1, 3\,--\,2, 5\,--\,4 and 6\,--\,5, which trace mostly the outer parts of
the $^{28}$SiO envelope.  The value \cite{GonzalezDelgado2003} derive for the $^{28}$SiO abundance is a factor of two 
lower than the $3.3 \times 10^{-5}$ abundance found by us. However, if the
lower mass-loss rate, $8\times10^{-8}$ and the smaller distance, 65 parsecs, considered by them
are taken into account the derived abundance should be even smaller, in the context of our model.
Our calculations overpredict the emission seen in these low-excitation transitions by a factor of four, 
consistent with the difference in the abundances that are obtained. \cite{Lucas1992} determined
the half-intensity angular radius for the $^{28}$SiO transition $J=2-1$ to be $0.9\pm 0.1$ arcseconds.
This value is substantially smaller than the value derived by \cite{GonzalezDelgado2003} for the $e$-folding radius
of the $^{28}$SiO abundance,
$2.4 \times 10^{15}$ cm or 2 arcseconds in the context of our model. The inconsistency is only apparent, as \citeauthor{GonzalezDelgado2003}
point out that these two radii are indeed expected to differ. The authors find that the $e$-folding radii determined by them are
about three times larger than the half-intensity radius of the $J=2-1$ $^{28}$SiO transition for model envelopes.
\cite{Schoier2004} observed the $^{28}$SiO $J=2-1$ transition of R\,Dor and L$^2$\,Pup with the Australia Telescope Compact Array.
By modelling the interferometric data, the authors found that the $^{28}$SiO abundance is better described by a two-component profile, a high abundance ($f_{\rm SiO} \approx 4 \times 10^{-5}$)
inner component and a lower abundance ($f_{\rm SiO} \approx 2-3 \times 10^{-6}$) extended component. The radius where the abundance drops is found to be
between 1 and $3 \times 10^{15}$\,cm.

The low-excitation transitions probe mostly the outer envelope, where dissociation occurs. The population of level $J=6$ of $^{28}$SiO reaches its maximum at
60 R$_\star$ (1.5 arcseconds or $1.8 \times 10^{15}$\,cm in the context of our model). Therefore, the abundances derived based on transitions from $J=6$ and
lower levels will depend on the assumed dissociation profile.
We have calculated models with smaller dissociation
radii and shallower dissociation profiles. We have done so by decreasing, respectively, the value of the
$e$-folding radius $r_{\rm e}$ and of the exponent $p$, initially kept at $p= 2$, in the expression for the $^{28}$SiO abundance profile (see Equation \ref{equ:diss_SiO}).
For reasonable values of these two parameters, the models are still unable to fit the low-excitation transitions.
An alternative possibility to explain simultaneously the high- and low-excitation lines
may be that $^{28}$SiO suffers from further depletion from the gas-phase in-between the region
where the lines observed by {\it Herschel} and those observed by SEST are excited.
At such large distances from the star, however, condensation and dissociation are indistinguishable on the basis of $^{28}$SiO line emission modelling.
The $J_{\rm up} > 10$ lines modelled by us, are all produced closer to the star and, therefore, trace the depletion of $^{28}$SiO independently of dissociation.
Thanks to the apparent complex nature of the $^{28}$SiO dissociation region and since the choice of dissociation radius does not have a significant impact
on the derived value for the $^{28}$SiO abundance, we do not attempt to fit the low-excitation $^{28}$SiO lines in detail. The value for the $^{28}$SiO depletion
obtained by us is representative of the inner wind, for $r < 1.5 \times 10^{15}$\,cm.

\section{Summary}
\label{sec:summary}

We present an analysis of the ortho-H$_2^{16}$O, para-H$_2^{16}$O, ortho-H$_2^{17}$O, para-H$_2^{17}$O, 
ortho-H$_2^{18}$O, para-H$_2^{18}$O and $^{28}$Si$^{16}$O emission from the wind of the nearby oxygen-rich 
AGB star W\,Hya, as measured by the three instruments on board {\it Herschel}.  The work builds on the
structure model of \cite{Khouri2014}, derived on the basis of $^{12}$CO lines, and is the first combined $^{12}$CO, H$_2^{16}$O and $^{28}$SiO
analysis of this source.

The original structure model poorly constrained the turbulence component of the velocity field in the outflow.
H$_2^{16}$O and $^{28}$SiO lines put much firmer constraints on the value of the turbulence velocity, essentially because
they are much optically thicker than the $^{12}$CO lines.  The presence of turbulence motions causes 
the H$_{2}$O and $^{28}$SiO lines to shift to longer wavelengths, which, when compared to $^{12}$CO profiles that form in
roughly the same part of the wind appear to imply a blue-wing absorption.
We find slightly different values for $\upsilon_{\rm turb}$ for ortho-H$_2^{16}$O and para-H$_2^{16}$O,
0.8 and 0.6 km\,s$^{-1}$ respectively, but as our code is not able to calculate models with a gradient in $\upsilon_{\rm turb}$, we have not explored
this further.

The abundance of ortho-H$_2^{16}$O and para-H$_2^{16}$O relative to H$_{2}$ are $(6^{+3}_{-2})\times 10^{-4}$ and
 $(3^{+2}_{-1})\times 10^{-4}$.  We also place constraints on the abundances of ortho-H$_2^{17}$O and para-H$_2^{17}$O,
and find an ortho-to-para ratio of 
2.5\,$^{+2.5}_{-1.0}$ --  in agreement with the value of three expected for AGB stars.
The $^{16}$O/$^{17}$O ratio is found to be 1250\,$^{+750}_{-450}$ and suggests that W\,Hya has an initial mass 
of about 1.5 M$_\odot$.
We find an $^{16}$O/$^{18}$O ratio of 190\,$^{+210}_{-90}$, which cannot be explained by the current generation
of evolutionary models. It might be reconciled with predictions if W\,Hya is richer in metals than
the Sun, but no firm conclusions can be drawn on this matter given the large uncertainties on the abundance
determination.

We find an $^{28}$SiO abundance between 10 and 100 R$_\star$ of $3.3 \pm 0.8 \times 10^{-5}$ relative to H$_2$.
Adding to this gas-phase abundance the abundance needed by our dust model, equivalent to $1.6 \times 10^{-5}$, 
we can account for all silicon in the wind of W\,Hya if a solar silicon-to-carbon ratio is assumed.
 
\begin{acknowledgements}
HIFI has been designed and built by a consortium of
institutes and university departments from across Europe, Canada, and the
United States under the leadership of SRON Netherlands Institute for Space
Research, Groningen, The Netherlands and with major contributions from
Germany, France, and the US. Consortium members are Canada: CSA,
U. Waterloo; France: CESR, LAB, LERMA, IRAM; Germany: KOSMA,
MPIfR, MPS; Ireland, NUI Maynooth; Italy: ASI, IFSI-INAF, Osservatorio
Astrofisico di Arcetri-INAF; Netherlands: SRON, TUD; Poland: CAMK, CBK;
Spain: Observatorio Astron\'omico Nacional (IGN), Centro de Astrobiolog\'{i}a
(CSIC-INTA). Sweden: Chalmers University of Technology Ð MC2, RSS \&
GARD; Onsala Space Observatory; Swedish National Space Board, Stockholm
University Ð SStockholm Observatory; Switzerland: ETH Zurich, FHNW; USA:
Caltech, JPL, NHSC.
PACS has been developed by a consortium of institutes
led by MPE (Germany) and including UVIE (Austria); KUL, CSL,
IMEC (Belgium); CEA, OAMP (France); MPIA (Germany); IFSI, OAP/AOT,
OAA/CAISMI, LENS, SISSA (Italy); IAC (Spain). This development has been
supported by the funding agencies BMVIT (Austria), ESA-PRODEX (Belgium),
CEA/CNES (France), DLR (Germany), ASI (Italy), and CICYT/MCYT (Spain).
SPIRE has been developed by a consortium of institutes led by Cardiff Univ. (UK) and
including Univ. Lethbridge (Canada); NAOC (China); CEA, LAM (France);
IFSI, Univ. Padua (Italy); IAC (Spain); Stockholm Observatory (Sweden);
Imperial College London, RAL, UCL-MSSL, UKATC, Univ. Sussex (UK);
Caltech, JPL, NHSC, Univ. Colorado (USA). This development has been supported
by national funding agencies: CSA (Canada); NAOC (China); CEA,
CNES, CNRS (France); ASI (Italy); MCINN (Spain); SNSB (Sweden); STFC
(UK); and NASA (USA).
HIFISTARS: {\it The physical and chemical properties
of circumstellar environments around evolved stars}, (P.I. V. Bujarrabal, is
a {\em Herschel}/HIFI guaranteed time key program (KPGT\_vbujarra\_1) devoted
to the study of the warm gas and water vapour contents of the molecular
envelopes around evolved stars: AGB stars, red super- and hyper-giants;
and their descendants: pre-planetary nebulae, planetary nebulae, and yellow
hyper-giants. HIFISTARS comprises 366 observations, totalling 11 186min
of {\em Herschel}/HIFI telescope time. See http://hifistars.oan.es; and
Key\_Programmes.shtml and UserProvidedDataProducts.shtml in the
Herschel web portal (http://herschel.esac.esa.int/) for additional details.
T.Kh. gratefully acknowledges the support from NWO grant 614.000.903.
F.K. is supported by the FWF project P23586 and the ffg ASAP project HIL.
MSc and RSz acknowledge support by the National
Science Center under grant (N 203 581040).
\end{acknowledgements}

\bibliographystyle{aa}
\bibliography{../../bibliography_2}

\begin{thebibliography}{71}
\expandafter\ifx\csname natexlab\endcsname\relax\def\natexlab#1{#1}\fi

\bibitem[{{Barlow} {et~al.}(1996){Barlow}, {Nguyen-Q-Rieu}, {Truong-Bach},
  {Cernicharo}, {Gonzalez-Alfonso}, {Liu}, {Cox}, {Sylvester}, {Clegg},
  {Griffin}, {Swinyard}, {Unger}, {Baluteau}, {Caux}, {Cohen}, {Cohen},
  {Emery}, {Fischer}, {Furniss}, {Glencross}, {Greenhouse}, {Gry}, {Joubert},
  {Lim}, {Lorenzetti}, {Nisini}, {Omont}, {Orfei}, {Pequignot}, {Saraceno},
  {Serra}, {Skinner}, {Smith}, {Walker}, {Armand}, {Burgdorf}, {Ewart}, {di
  Giorgio}, {Molinari}, {Price}, {Sidher}, {Texier}, \& {Trams}}]{Barlow1996}
{Barlow}, M.~J., {Nguyen-Q-Rieu}, {Truong-Bach}, {et~al.} 1996, \aap, 315, L241

\bibitem[{{Begemann} {et~al.}(1997){Begemann}, {Dorschner}, {Henning},
  {Mutschke}, {Guertler}, {Koempe}, \& {Nass}}]{Begemann1997}
{Begemann}, B., {Dorschner}, J., {Henning}, T., {et~al.} 1997, \apj, 476, 199

\bibitem[{{Bieging} {et~al.}(2000){Bieging}, {Shaked}, \&
  {Gensheimer}}]{Bieging2000}
{Bieging}, J.~H., {Shaked}, S., \& {Gensheimer}, P.~D. 2000, \apj, 543, 897

\bibitem[{{Bladh} \& {H{\"o}fner}(2012)}]{Bladh2012}
{Bladh}, S. \& {H{\"o}fner}, S. 2012, \aap, 546, A76

\bibitem[{{Boothroyd} {et~al.}(1994){Boothroyd}, {Sackmann}, \&
  {Wasserburg}}]{Boothroyd1994}
{Boothroyd}, A.~I., {Sackmann}, I.-J., \& {Wasserburg}, G.~J. 1994, \apjl, 430,
  L77

\bibitem[{{Bujarrabal} {et~al.}(1986){Bujarrabal}, {Planesas},
  {Martin-Pintado}, {Gomez-Gonzalez}, \& {del Romero}}]{Bujarrabal1986}
{Bujarrabal}, V., {Planesas}, P., {Martin-Pintado}, J., {Gomez-Gonzalez}, J.,
  \& {del Romero}, A. 1986, \aap, 162, 157

\bibitem[{{Busso} {et~al.}(1999){Busso}, {Gallino}, \&
  {Wasserburg}}]{Busso1999}
{Busso}, M., {Gallino}, R., \& {Wasserburg}, G.~J. 1999, \araa, 37, 239

\bibitem[{{Charbonnel} \& {Lagarde}(2010)}]{Charbonnel2010}
{Charbonnel}, C. \& {Lagarde}, N. 2010, \aap, 522, A10

\bibitem[{{Cherchneff}(2006)}]{Cherchneff2006}
{Cherchneff}, I. 2006, \aap, 456, 1001

\bibitem[{{Cox} {et~al.}(2012){Cox}, {Kerschbaum}, {van Marle}, {Decin},
  {Ladjal}, {Mayer}, {Groenewegen}, {van Eck}, {Royer}, {Ottensamer}, {Ueta},
  {Jorissen}, {Mecina}, {Meliani}, {Luntzer}, {Blommaert}, {Posch},
  {Vandenbussche}, \& {Waelkens}}]{Cox2012}
{Cox}, N.~L.~J., {Kerschbaum}, F., {van Marle}, A.-J., {et~al.} 2012, \aap,
  537, A35

\bibitem[{{de Graauw} {et~al.}(2010){de Graauw}, {Helmich}, {Phillips},
  {Stutzki}, {Caux}, {Whyborn}, {Dieleman}, {Roelfsema}, {Aarts}, {Assendorp},
  {Bachiller}, {Baechtold}, {Barcia}, {Beintema}, {Belitsky}, {Benz}, {Bieber},
  {Boogert}, {Borys}, {Bumble}, {Ca{\"i}s}, {Caris}, {Cerulli-Irelli},
  {Chattopadhyay}, {Cherednichenko}, {Ciechanowicz}, {Coeur-Joly}, {Comito},
  {Cros}, {de Jonge}, {de Lange}, {Delforges}, {Delorme}, {den Boggende},
  {Desbat}, {Diez-Gonz{\'a}lez}, {di Giorgio}, {Dubbeldam}, {Edwards},
  {Eggens}, {Erickson}, {Evers}, {Fich}, {Finn}, {Franke}, {Gaier}, {Gal},
  {Gao}, {Gallego}, {Gauffre}, {Gill}, {Glenz}, {Golstein}, {Goulooze},
  {Gunsing}, {G{\"u}sten}, {Hartogh}, {Hatch}, {Higgins}, {Honingh}, {Huisman},
  {Jackson}, {Jacobs}, {Jacobs}, {Jarchow}, {Javadi}, {Jellema}, {Justen},
  {Karpov}, {Kasemann}, {Kawamura}, {Keizer}, {Kester}, {Klapwijk}, {Klein},
  {Kollberg}, {Kooi}, {Kooiman}, {Kopf}, {Krause}, {Krieg}, {Kramer},
  {Kruizenga}, {Kuhn}, {Laauwen}, {Lai}, {Larsson}, {Leduc}, {Leinz}, {Lin},
  {Liseau}, {Liu}, {Loose}, {L{\'o}pez-Fernandez}, {Lord}, {Luinge}, {Marston},
  {Mart{\'{\i}}n-Pintado}, {Maestrini}, {Maiwald}, {McCoey}, {Mehdi}, {Megej},
  {Melchior}, {Meinsma}, {Merkel}, {Michalska}, {Monstein}, {Moratschke},
  {Morris}, {Muller}, {Murphy}, {Naber}, {Natale}, {Nowosielski}, {Nuzzolo},
  {Olberg}, {Olbrich}, {Orfei}, {Orleanski}, {Ossenkopf}, {Peacock}, {Pearson},
  {Peron}, {Phillip-May}, {Piazzo}, {Planesas}, {Rataj}, {Ravera}, {Risacher},
  {Salez}, {Samoska}, {Saraceno}, {Schieder}, {Schlecht}, {Schl{\"o}der},
  {Schm{\"u}lling}, {Schultz}, {Schuster}, {Siebertz}, {Smit}, {Szczerba},
  {Shipman}, {Steinmetz}, {Stern}, {Stokroos}, {Teipen}, {Teyssier}, {Tils},
  {Trappe}, {van Baaren}, {van Leeuwen}, {van de Stadt}, {Visser}, {Wildeman},
  {Wafelbakker}, {Ward}, {Wesselius}, {Wild}, {Wulff}, {Wunsch}, {Tielens},
  {Zaal}, {Zirath}, {Zmuidzinas}, \& {Zwart}}]{deGraauw2010}
{de Graauw}, T., {Helmich}, F.~P., {Phillips}, T.~G., {et~al.} 2010, \aap, 518,
  L6

\bibitem[{{Decin} {et~al.}(2010{\natexlab{a}}){Decin}, {De Beck},
  {Br{\"u}nken}, {M{\"u}ller}, {Menten}, {Kim}, {Willacy}, {de Koter}, \&
  {Wyrowski}}]{Decin2010a}
{Decin}, L., {De Beck}, E., {Br{\"u}nken}, S., {et~al.} 2010{\natexlab{a}},
  \aap, 516, A69

\bibitem[{{Decin} {et~al.}(2006){Decin}, {Hony}, {de Koter}, {Justtanont},
  {Tielens}, \& {Waters}}]{Decin2006}
{Decin}, L., {Hony}, S., {de Koter}, A., {et~al.} 2006, \aap, 456, 549

\bibitem[{{Decin} {et~al.}(2010{\natexlab{b}}){Decin}, {Justtanont}, {De Beck},
  {Lombaert}, {de Koter}, {Waters}, {Marston}, {Teyssier}, {Sch{\"o}ier},
  {Bujarrabal}, {Alcolea}, {Cernicharo}, {Dominik}, {Melnick}, {Menten},
  {Neufeld}, {Olofsson}, {Planesas}, {Schmidt}, {Szczerba}, {de Graauw},
  {Helmich}, {Roelfsema}, {Dieleman}, {Morris}, {Gallego},
  {D{\'{\i}}ez-Gonz{\'a}lez}, \& {Caux}}]{Decin2010}
{Decin}, L., {Justtanont}, K., {De Beck}, E., {et~al.} 2010{\natexlab{b}},
  \aap, 521, L4

\bibitem[{{El Eid}(1994)}]{ElEid1994}
{El Eid}, M.~F. 1994, \aap, 285, 915

\bibitem[{{Faure} {et~al.}(2007){Faure}, {Crimier}, {Ceccarelli}, {Valiron},
  {Wiesenfeld}, \& {Dubernet}}]{Faure2007}
{Faure}, A., {Crimier}, N., {Ceccarelli}, C., {et~al.} 2007, \aap, 472, 1029

\bibitem[{{Gail} \& {Sedlmayr}(1999)}]{Gail1999}
{Gail}, H.-P. \& {Sedlmayr}, E. 1999, \aap, 347, 594

\bibitem[{{Gonz{\'a}lez Delgado} {et~al.}(2003){Gonz{\'a}lez Delgado},
  {Olofsson}, {Kerschbaum}, {Sch{\"o}ier}, {Lindqvist}, \&
  {Groenewegen}}]{GonzalezDelgado2003}
{Gonz{\'a}lez Delgado}, D., {Olofsson}, H., {Kerschbaum}, F., {et~al.} 2003,
  \aap, 411, 123

\bibitem[{{Griffin} {et~al.}(2010){Griffin}, {Abergel}, {Abreu}, {Ade},
  {Andr{\'e}}, {Augueres}, {Babbedge}, {Bae}, {Baillie}, {Baluteau}, {Barlow},
  {Bendo}, {Benielli}, {Bock}, {Bonhomme}, {Brisbin}, {Brockley-Blatt},
  {Caldwell}, {Cara}, {Castro-Rodriguez}, {Cerulli}, {Chanial}, {Chen},
  {Clark}, {Clements}, {Clerc}, {Coker}, {Communal}, {Conversi}, {Cox},
  {Crumb}, {Cunningham}, {Daly}, {Davis}, {de Antoni}, {Delderfield}, {Devin},
  {di Giorgio}, {Didschuns}, {Dohlen}, {Donati}, {Dowell}, {Dowell}, {Duband},
  {Dumaye}, {Emery}, {Ferlet}, {Ferrand}, {Fontignie}, {Fox}, {Franceschini},
  {Frerking}, {Fulton}, {Garcia}, {Gastaud}, {Gear}, {Glenn}, {Goizel},
  {Griffin}, {Grundy}, {Guest}, {Guillemet}, {Hargrave}, {Harwit}, {Hastings},
  {Hatziminaoglou}, {Herman}, {Hinde}, {Hristov}, {Huang}, {Imhof}, {Isaak},
  {Israelsson}, {Ivison}, {Jennings}, {Kiernan}, {King}, {Lange}, {Latter},
  {Laurent}, {Laurent}, {Leeks}, {Lellouch}, {Levenson}, {Li}, {Li},
  {Lilienthal}, {Lim}, {Liu}, {Lu}, {Madden}, {Mainetti}, {Marliani}, {McKay},
  {Mercier}, {Molinari}, {Morris}, {Moseley}, {Mulder}, {Mur}, {Naylor},
  {Nguyen}, {O'Halloran}, {Oliver}, {Olofsson}, {Olofsson}, {Orfei}, {Page},
  {Pain}, {Panuzzo}, {Papageorgiou}, {Parks}, {Parr-Burman}, {Pearce},
  {Pearson}, {P{\'e}rez-Fournon}, {Pinsard}, {Pisano}, {Podosek}, {Pohlen},
  {Polehampton}, {Pouliquen}, {Rigopoulou}, {Rizzo}, {Roseboom}, {Roussel},
  {Rowan-Robinson}, {Rownd}, {Saraceno}, {Sauvage}, {Savage}, {Savini},
  {Sawyer}, {Scharmberg}, {Schmitt}, {Schneider}, {Schulz}, {Schwartz},
  {Shafer}, {Shupe}, {Sibthorpe}, {Sidher}, {Smith}, {Smith}, {Smith},
  {Spencer}, {Stobie}, {Sudiwala}, {Sukhatme}, {Surace}, {Stevens}, {Swinyard},
  {Trichas}, {Tourette}, {Triou}, {Tseng}, {Tucker}, {Turner}, {Vaccari},
  {Valtchanov}, {Vigroux}, {Virique}, {Voellmer}, {Walker}, {Ward}, {Waskett},
  {Weilert}, {Wesson}, {White}, {Whitehouse}, {Wilson}, {Winter}, {Woodcraft},
  {Wright}, {Xu}, {Zavagno}, {Zemcov}, {Zhang}, \& {Zonca}}]{Griffin2010}
{Griffin}, M.~J., {Abergel}, A., {Abreu}, A., {et~al.} 2010, \aap, 518, L3

\bibitem[{{Groenewegen}(1994)}]{Groenewegen1994}
{Groenewegen}, M.~A.~T. 1994, \aap, 290, 531

\bibitem[{{Groenewegen} {et~al.}(2011){Groenewegen}, {Waelkens}, {Barlow},
  {Kerschbaum}, {Garcia-Lario}, {Cernicharo}, {Blommaert}, {Bouwman}, {Cohen},
  {Cox}, {Decin}, {Exter}, {Gear}, {Gomez}, {Hargrave}, {Henning},
  {Hutsem{\'e}kers}, {Ivison}, {Jorissen}, {Krause}, {Ladjal}, {Leeks}, {Lim},
  {Matsuura}, {Naz{\'e}}, {Olofsson}, {Ottensamer}, {Polehampton}, {Posch},
  {Rauw}, {Royer}, {Sibthorpe}, {Swinyard}, {Ueta}, {Vamvatira-Nakou},
  {Vandenbussche}, {van de Steene}, {van Eck}, {van Hoof}, {van Winckel},
  {Verdugo}, \& {Wesson}}]{Groenewegen2011}
{Groenewegen}, M.~A.~T., {Waelkens}, C., {Barlow}, M.~J., {et~al.} 2011, \aap,
  526, A162

\bibitem[{{Habing} \& {Olofsson}(2003)}]{Habing2003}
{Habing}, H.~J. \& {Olofsson}, H., eds. 2003, {Asymptotic Giant Branch Stars}

\bibitem[{{Harwit} \& {Bergin}(2002)}]{Harwit2002}
{Harwit}, M. \& {Bergin}, E.~A. 2002, \apjl, 565, L105

\bibitem[{{Hawkins}(1990)}]{Hawkins1990}
{Hawkins}, G.~W. 1990, \aap, 229, L5

\bibitem[{{Henning} {et~al.}(1995){Henning}, {Begemann}, {Mutschke}, \&
  {Dorschner}}]{Henning1995}
{Henning}, T., {Begemann}, B., {Mutschke}, H., \& {Dorschner}, J. 1995, \aaps,
  112, 143

\bibitem[{{H{\"o}fner}(2008)}]{Hofner2008}
{H{\"o}fner}, S. 2008, \aap, 491, L1

\bibitem[{{Huggins} \& {Healy}(1986)}]{Huggins1986}
{Huggins}, P.~J. \& {Healy}, A.~P. 1986, \apj, 304, 418

\bibitem[{{Iben}(1975)}]{Iben1975}
{Iben}, Jr., I. 1975, \apj, 196, 525

\bibitem[{{Iben} \& {Renzini}(1983)}]{Iben1983}
{Iben}, Jr., I. \& {Renzini}, A. 1983, \araa, 21, 271

\bibitem[{{Imai} {et~al.}(2010){Imai}, {Nakashima}, {Deguchi}, {Yamauchi},
  {Nakagawa}, \& {Nagayama}}]{Imai2010}
{Imai}, H., {Nakashima}, J.-I., {Deguchi}, S., {et~al.} 2010, \pasj, 62, 431

\bibitem[{{Justtanont} {et~al.}(2005){Justtanont}, {Bergman}, {Larsson},
  {Olofsson}, {Sch{\"o}ier}, {Frisk}, {Hasegawa}, {Hjalmarson}, {Kwok},
  {Olberg}, {Sandqvist}, {Volk}, \& {Elitzur}}]{Justtanont2005}
{Justtanont}, K., {Bergman}, P., {Larsson}, B., {et~al.} 2005, \aap, 439, 627

\bibitem[{{Justtanont} {et~al.}(2004){Justtanont}, {de Jong}, {Tielens},
  {Feuchtgruber}, \& {Waters}}]{Justtanont2004}
{Justtanont}, K., {de Jong}, T., {Tielens}, A.~G.~G.~M., {Feuchtgruber}, H., \&
  {Waters}, L.~B.~F.~M. 2004, \aap, 417, 625

\bibitem[{{Justtanont} {et~al.}(2012){Justtanont}, {Khouri}, {Maercker},
  {Alcolea}, {Decin}, {Olofsson}, {Sch{\"o}ier}, {Bujarrabal}, {Marston},
  {Teyssier}, {Cernicharo}, {Dominik}, {de Koter}, {Melnick}, {Menten},
  {Neufeld}, {Planesas}, {Schmidt}, {Szczerba}, \& {Waters}}]{Justtanont2012}
{Justtanont}, K., {Khouri}, T., {Maercker}, M., {et~al.} 2012, \aap, 537, A144

\bibitem[{{Justtanont} {et~al.}(2013){Justtanont}, {Teyssier}, {Barlow},
  {Matsuura}, {Swinyard}, {Waters}, \& {Yates}}]{Justtanont2013}
{Justtanont}, K., {Teyssier}, D., {Barlow}, M.~J., {et~al.} 2013, \aap, 556,
  A101

\bibitem[{{Justtanont} \& {Tielens}(1992)}]{Justtanont1992}
{Justtanont}, K. \& {Tielens}, A.~G.~G.~M. 1992, \apj, 389, 400

\bibitem[{{Karakas}(2010)}]{Karakas2010}
{Karakas}, A.~I. 2010, \mnras, 403, 1413

\bibitem[{{Karakas}(2011)}]{Karakas2011}
{Karakas}, A.~I. 2011, in Astronomical Society of the Pacific Conference
  Series, Vol. 445, Why Galaxies Care about AGB Stars II: Shining Examples and
  Common Inhabitants, ed. F.~{Kerschbaum}, T.~{Lebzelter}, \& R.~F. {Wing}, 3

\bibitem[{{Kessler} {et~al.}(1996){Kessler}, {Steinz}, {Anderegg}, {Clavel},
  {Drechsel}, {Estaria}, {Faelker}, {Riedinger}, {Robson}, {Taylor}, \&
  {Xim{\'e}nez de Ferr{\'a}n}}]{Kessler1996}
{Kessler}, M.~F., {Steinz}, J.~A., {Anderegg}, M.~E., {et~al.} 1996, \aap, 315,
  L27

\bibitem[{{Khouri} {et~al.}(2014){Khouri}, {de Koter}, {Decin}, {Waters},
  {Lombaert}, {Royer}, {Swinyard}, {Barlow}, {Alcolea}, {Blommaert},
  {Bujarrabal}, {Cernicharo}, {Groenewegen}, {Justtanont}, {Kerschbaum},
  {Maercker}, {Marston}, {Matsuura}, {Melnick}, {Menten}, {Olofsson},
  {Planesas}, {Polehampton}, {Posch}, {Schmidt}, {Szczerba}, {Vandenbussche},
  \& {Yates}}]{Khouri2014}
{Khouri}, T., {de Koter}, A., {Decin}, L., {et~al.} 2014, \aap, 561, A5 (Paper
  I)

\bibitem[{{Knapp} {et~al.}(2003){Knapp}, {Pourbaix}, {Platais}, \&
  {Jorissen}}]{Knapp2003}
{Knapp}, G.~R., {Pourbaix}, D., {Platais}, I., \& {Jorissen}, A. 2003, \aap,
  403, 993

\bibitem[{{Landre} {et~al.}(1990){Landre}, {Prantzos}, {Aguer}, {Bogaert},
  {Lefebvre}, \& {Thibaud}}]{Landre1990}
{Landre}, V., {Prantzos}, N., {Aguer}, P., {et~al.} 1990, \aap, 240, 85

\bibitem[{{Langhoff} \& {Bauschlicher}(1993)}]{Langhoff1993}
{Langhoff}, S.~R. \& {Bauschlicher}, Jr., C.~W. 1993, Chemical Physics Letters,
  211, 305

\bibitem[{{Lattanzio} \& {Boothroyd}(1997)}]{Lattanzio1997}
{Lattanzio}, J.~C. \& {Boothroyd}, A.~I. 1997, in American Institute of Physics
  Conference Series, Vol. 402, American Institute of Physics Conference Series,
  ed. T.~J. {Bernatowicz} \& E.~{Zinner}, 85--114

\bibitem[{{Lombaert} {et~al.}(2013){Lombaert}, {Decin}, {de Koter},
  {Blommaert}, {Royer}, {De Beck}, {de Vries}, {Khouri}, \&
  {Min}}]{Lombaert2013}
{Lombaert}, R., {Decin}, L., {de Koter}, A., {et~al.} 2013, \aap, 554, A142

\bibitem[{{Lucas} {et~al.}(1992){Lucas}, {Bujarrabal}, {Guilloteau},
  {Bachiller}, {Baudry}, {Cernicharo}, {Delannoy}, {Forveille}, {Guelin}, \&
  {Radford}}]{Lucas1992}
{Lucas}, R., {Bujarrabal}, V., {Guilloteau}, S., {et~al.} 1992, \aap, 262, 491

\bibitem[{{Maercker} {et~al.}(2009){Maercker}, {Sch{\"o}ier}, {Olofsson},
  {Bergman}, {Frisk}, {.~Hjalmarson}, {Justtanont}, {Kwok}, {Larsson},
  {Olberg}, \& {Sandqvist}}]{Maercker2009}
{Maercker}, M., {Sch{\"o}ier}, F.~L., {Olofsson}, H., {et~al.} 2009, \aap, 494,
  243

\bibitem[{{Maercker} {et~al.}(2008){Maercker}, {Sch{\"o}ier}, {Olofsson},
  {Bergman}, \& {Ramstedt}}]{Maercker2008}
{Maercker}, M., {Sch{\"o}ier}, F.~L., {Olofsson}, H., {Bergman}, P., \&
  {Ramstedt}, S. 2008, \aap, 479, 779

\bibitem[{{Mamon} {et~al.}(1988){Mamon}, {Glassgold}, \& {Huggins}}]{Mamon1988}
{Mamon}, G.~A., {Glassgold}, A.~E., \& {Huggins}, P.~J. 1988, \apj, 328, 797

\bibitem[{{Matsuura} {et~al.}(2013){Matsuura}, {Yates}, {Barlow}, {Swinyard},
  {Royer}, {Cernicharo}, {Decin}, {Wesson}, {Polehampton}, {Blommaert},
  {Groenewegen}, {Van de Steene}, \& {van Hoof}}]{Matsuura2013}
{Matsuura}, M., {Yates}, J.~A., {Barlow}, M.~J., {et~al.} 2013, \mnras

\bibitem[{{Melnick} {et~al.}(2000){Melnick}, {Stauffer}, {Ashby}, {Bergin},
  {Chin}, {Erickson}, {Goldsmith}, {Harwit}, {Howe}, {Kleiner}, {Koch},
  {Neufeld}, {Patten}, {Plume}, {Schieder}, {Snell}, {Tolls}, {Wang},
  {Winnewisser}, \& {Zhang}}]{Melnick2000}
{Melnick}, G.~J., {Stauffer}, J.~R., {Ashby}, M.~L.~N., {et~al.} 2000, \apjl,
  539, L77

\bibitem[{{Menten} {et~al.}(2010){Menten}, {Bujarrabal}, {Alcolea},
  {Cernicharo}, {Decin}, {Dominik}, {Justtanont}, {de Koter}, {Marston},
  {Melnick}, {Neufeld}, {Olofsson}, {Planesas}, {Pulecka}, {Schmidt},
  {Schoier}, {Szczerba}, {Teyssier}, \& {Waters}}]{Menten2010}
{Menten}, K., {Bujarrabal}, V., {Alcolea}, J., {et~al.} 2010, in COSPAR
  Meeting, Vol.~38, 38th COSPAR Scientific Assembly, 2490

\bibitem[{{Min} {et~al.}(2009){Min}, {Dullemond}, {Dominik}, {de Koter}, \&
  {Hovenier}}]{Min2009}
{Min}, M., {Dullemond}, C.~P., {Dominik}, C., {de Koter}, A., \& {Hovenier},
  J.~W. 2009, \aap, 497, 155

\bibitem[{{Netzer} \& {Knapp}(1987)}]{Netzer1987}
{Netzer}, N. \& {Knapp}, G.~R. 1987, \apj, 323, 734

\bibitem[{{Neufeld} {et~al.}(1996){Neufeld}, {Chen}, {Melnick}, {de Graauw},
  {Feuchtgruber}, {Haser}, {Lutz}, \& {Harwit}}]{Neufeld1996}
{Neufeld}, D.~A., {Chen}, W., {Melnick}, G.~J., {et~al.} 1996, \aap, 315, L237

\bibitem[{{Nordh} {et~al.}(2003){Nordh}, {von Sch{\'e}ele}, {Frisk}, {Ahola},
  {Booth}, {Encrenaz}, {Hjalmarson}, {Kendall}, {Kyr{\"o}l{\"a}}, {Kwok},
  {Lecacheux}, {Leppelmeier}, {Llewellyn}, {Mattila}, {M{\'e}gie}, {Murtagh},
  {Rougeron}, \& {Witt}}]{Nordh2003}
{Nordh}, H.~L., {von Sch{\'e}ele}, F., {Frisk}, U., {et~al.} 2003, \aap, 402,
  L21

\bibitem[{{Norris} {et~al.}(2012){Norris}, {Tuthill}, {Ireland}, {Lacour},
  {Zijlstra}, {Lykou}, {Evans}, {Stewart}, \& {Bedding}}]{Norris2012}
{Norris}, B.~R.~M., {Tuthill}, P.~G., {Ireland}, M.~J., {et~al.} 2012, \nat,
  484, 220

\bibitem[{{Palmerini} {et~al.}(2011){Palmerini}, {La Cognata}, {Cristallo}, \&
  {Busso}}]{Palmerini2011}
{Palmerini}, S., {La Cognata}, M., {Cristallo}, S., \& {Busso}, M. 2011, \apj,
  729, 3

\bibitem[{{Pilbratt} {et~al.}(2010){Pilbratt}, {Riedinger}, {Passvogel},
  {Crone}, {Doyle}, {Gageur}, {Heras}, {Jewell}, {Metcalfe}, {Ott}, \&
  {Schmidt}}]{Pilbratt2010}
{Pilbratt}, G.~L., {Riedinger}, J.~R., {Passvogel}, T., {et~al.} 2010, \aap,
  518, L1

\bibitem[{{Poglitsch} {et~al.}(2010){Poglitsch}, {Waelkens}, {Geis},
  {Feuchtgruber}, {Vandenbussche}, {Rodriguez}, {Krause}, {Renotte}, {van
  Hoof}, {Saraceno}, {Cepa}, {Kerschbaum}, {Agn{\`e}se}, {Ali}, {Altieri},
  {Andreani}, {Augueres}, {Balog}, {Barl}, {Bauer}, {Belbachir}, {Benedettini},
  {Billot}, {Boulade}, {Bischof}, {Blommaert}, {Callut}, {Cara}, {Cerulli},
  {Cesarsky}, {Contursi}, {Creten}, {De Meester}, {Doublier}, {Doumayrou},
  {Duband}, {Exter}, {Genzel}, {Gillis}, {Gr{\"o}zinger}, {Henning},
  {Herreros}, {Huygen}, {Inguscio}, {Jakob}, {Jamar}, {Jean}, {de Jong},
  {Katterloher}, {Kiss}, {Klaas}, {Lemke}, {Lutz}, {Madden}, {Marquet},
  {Martignac}, {Mazy}, {Merken}, {Montfort}, {Morbidelli}, {M{\"u}ller},
  {Nielbock}, {Okumura}, {Orfei}, {Ottensamer}, {Pezzuto}, {Popesso},
  {Putzeys}, {Regibo}, {Reveret}, {Royer}, {Sauvage}, {Schreiber}, {Stegmaier},
  {Schmitt}, {Schubert}, {Sturm}, {Thiel}, {Tofani}, {Vavrek}, {Wetzstein},
  {Wieprecht}, \& {Wiezorrek}}]{Poglitsch2010}
{Poglitsch}, A., {Waelkens}, C., {Geis}, N., {et~al.} 2010, \aap, 518, L2

\bibitem[{{Rothman} {et~al.}(2009){Rothman}, {Gordon}, {Barbe}, {Benner},
  {Bernath}, {Birk}, {Boudon}, {Brown}, {Campargue}, {Champion}, {Chance},
  {Coudert}, {Dana}, {Devi}, {Fally}, {Flaud}, {Gamache}, {Goldman},
  {Jacquemart}, {Kleiner}, {Lacome}, {Lafferty}, {Mandin}, {Massie},
  {Mikhailenko}, {Miller}, {Moazzen-Ahmadi}, {Naumenko}, {Nikitin}, {Orphal},
  {Perevalov}, {Perrin}, {Predoi-Cross}, {Rinsland}, {Rotger}, {{\v S}ime{\v
  c}kov{\'a}}, {Smith}, {Sung}, {Tashkun}, {Tennyson}, {Toth}, {Vandaele}, \&
  {Vander Auwera}}]{Rothman2009}
{Rothman}, L.~S., {Gordon}, I.~E., {Barbe}, A., {et~al.} 2009, \jqsrt, 110, 533

\bibitem[{{Sch{\"o}ier} {et~al.}(2004){Sch{\"o}ier}, {Olofsson}, {Wong},
  {Lindqvist}, \& {Kerschbaum}}]{Schoier2004}
{Sch{\"o}ier}, F.~L., {Olofsson}, H., {Wong}, T., {Lindqvist}, M., \&
  {Kerschbaum}, F. 2004, \aap, 422, 651

\bibitem[{{Sch{\"o}ier} {et~al.}(2005){Sch{\"o}ier}, {van der Tak}, {van
  Dishoeck}, \& {Black}}]{Schoier2005}
{Sch{\"o}ier}, F.~L., {van der Tak}, F.~F.~S., {van Dishoeck}, E.~F., \&
  {Black}, J.~H. 2005, \aap, 432, 369

\bibitem[{{Sharp} \& {Huebner}(1990)}]{Sharp1990}
{Sharp}, C.~M. \& {Huebner}, W.~F. 1990, \apjs, 72, 417

\bibitem[{{Stoesz} \& {Herwig}(2003)}]{Stoesz2003}
{Stoesz}, J.~A. \& {Herwig}, F. 2003, \mnras, 340, 763

\bibitem[{{Sylvester} {et~al.}(1997){Sylvester}, {Barlow}, {Nguyen-Q-Rieu},
  {Liu}, {Skinner}, {Cohen}, {Lim}, {Cox}, {Truong-Bach}, {Smith}, \&
  {Habing}}]{Sylvester1997}
{Sylvester}, R.~J., {Barlow}, M.~J., {Nguyen-Q-Rieu}, {et~al.} 1997, \mnras,
  291, L42

\bibitem[{{Timmes} {et~al.}(1995){Timmes}, {Woosley}, \& {Weaver}}]{Timmes1995}
{Timmes}, F.~X., {Woosley}, S.~E., \& {Weaver}, T.~A. 1995, \apjs, 98, 617

\bibitem[{{Vlemmings} {et~al.}(2011){Vlemmings}, {Humphreys}, \&
  {Franco-Hern{\'a}ndez}}]{Vlemmings2011}
{Vlemmings}, W.~H.~T., {Humphreys}, E.~M.~L., \& {Franco-Hern{\'a}ndez}, R.
  2011, \apj, 728, 149

\bibitem[{{Winters} {et~al.}(2000){Winters}, {Le Bertre}, {Jeong}, {Helling},
  \& {Sedlmayr}}]{Winters2000}
{Winters}, J.~M., {Le Bertre}, T., {Jeong}, K.~S., {Helling}, C., \&
  {Sedlmayr}, E. 2000, \aap, 361, 641

\bibitem[{{Woitke}(2006)}]{Woitke2006}
{Woitke}, P. 2006, \aap, 460, L9

\bibitem[{{Zhao-Geisler} {et~al.}(2011){Zhao-Geisler}, {Quirrenbach},
  {K{\"o}hler}, {Lopez}, \& {Leinert}}]{Zhao-Geisler2011}
{Zhao-Geisler}, R., {Quirrenbach}, A., {K{\"o}hler}, R., {Lopez}, B., \&
  {Leinert}, C. 2011, \aap, 530, A120

\bibitem[{{Zubko} \& {Elitzur}(2000)}]{Zubko2000}
{Zubko}, V. \& {Elitzur}, M. 2000, \apjl, 544, L137

\end{thebibliography}

\begin{appendix}

\section{Molecular models}
\label{sec:AppendixA}

When modelling the H$_2^{16}$O transitions for all isotopologues, we include the 45 lowest levels of the ground and first vibrational states (i.e.
the bending mode $\nu_2 = 1$ at 6.3 $\mu$m). For the two spin isomers of the main isotopologue, we have also included excitation to the first excited vibrational
state of the asymmetric stretching mode ($\nu_3 = 1$). The difference on the model line fluxes due to the inclusion of the $\nu_3 = 1$ level is found to be 20\% at maximum
\citep{Decin2010a}. The frequencies, level energies and Einstein A coefficients were retrieved from the HITRAN H$_2^{16}$O line list \citep{Rothman2009}. The collisional
rates between H$_2^{16}$O and H$_2$ were extracted from \cite{Faure2007}.

Following \cite{Decin2010a}, we consider the 40 lowest rotational levels of the ground and first vibrationally excited states when modelling the $^{28}$SiO transitions.
The Einstein A coefficients, energy levels and frequencies were taken from \cite{Langhoff1993}. The collisional rates between $^{28}$SiO and H$_2$ were retrieved from the LAMBDA-database \citep{Schoier2005}.

\section{Observed $^{28}$SiO and H$_2^{16}$O line fluxes}
\label{sec:AppendixB}

\begin{table}[h]
\caption{Integrated line fluxes and uncertainties for the $^{28}$SiO transitions observed with {\it Herschel} for 
              W\,Hya. The column $Instr.$ lists the spectrograph used: HIFI, PACS, or SPIRE. }
\label{tab:obs_SiO}
\centering
\begin{tabular}{ c  r c  r  c   }
\hline

$J_{\rm up}$ & $\nu_0$ & $Instr.$ & $E $ & Flux   \\ 
                      & [GHz] & &  [K] & [$10^{-17}$ W\,m$^{-2}$] \\
\hline\\[-8pt]
11 & 477.50 & S & 137.5     & $3.5 \pm 0.7$ \\
12 & 520.88 & S & 162.5     & $2.2 \pm 0.5$ \\
13 & 564.25 & S & 189.6     & $2.5 \pm 0.5$ \\
14 & 607.61 & S & 218.8     & $2.0 \pm 0.5$ \\
14 & 607.61 & H & 218.8     & $2.1 \pm 0.4$ \\
15 & 650.96 & S & 250.0     & $3.1 \pm 0.7$  \\
16 & 694.29 & S & 283.3     & $2.4 \pm 0.5$ \\
16 & 694.29 & H & 283.3     & $2.5 \pm 0.5$ \\
17 & 737.62 & S & 318.7     & $2.1 \pm 0.5$ \\
18 & 780.93 & S & 356.2     & $3.5 \pm 0.7$ \\
19 & 824.24 & S & 395.8     & $3.2 \pm 0.7$ \\
20 & 867.52 & S & 437.4     & $3.0 \pm 0.7$ \\
21 & 910.80 & S & 481.1     & $3.7 \pm 0.8$ \\
22 & 954.05 & S & 526.9     & $4.5 \pm 0.9$ \\
23 & 997.30 & S & 574.8     & $3.4 \pm 0.8$ \\
24 & 1040.52 & S & 624.7   & $3.4 \pm 0.8$ \\
25 & 1083.73 & S & 676.7   & $3.2 \pm 0.7$ \\
26 & 1126.92 & S & 730.8   & $4.8 \pm 1.0$ \\
27 & 1170.09 & S & 787.0   & $2.8 \pm 0.7$ \\
28 & 1213.25 & S & 845.2   & $4.6 \pm 1.0$  \\
29 & 1256.38 & S & 905.5   & $4.3 \pm 0.9$  \\
31 & 1342.58 & S & 1032.3 & $4.0 \pm 0.9$  \\
32 & 1385.65 & S & 1098.8 & $3.0 \pm 0.9$ \\
33 & 1428.69 & S & 1167.4 & $5.7 \pm 1.8$ \\
34 & 1471.72 & S & 1238.0 & $4.2 \pm 1.0$ \\
35 & 1514.71 & S & 1310.7 & $3.0 \pm 0.9$ \\
37 & 1600.63 & P & 1462.3 & $5.1 \pm 1.5$ \\
\hline

\end{tabular}
\end{table}

\begin{center}
\onecolumn
\LTcapwidth=\textwidth
{\footnotesize
\begin{longtable}{l @{}r r c l c c c c c c c }
\caption{Extracted ortho-H$_2^{16}$O lines from the PACS observations. In the fifth column, we indicate if the line was flagged as a blend. We have specified if the blend happens with a known
ortho- or para-H$_2^{16}$O transition using, respectively, the superscripts $^{\rm o}$ and $^{\rm p}$. Lines that were excluded due to masering happening in any of the models
are identified by the superscript $^{\rm m}$.}\\
\label{table:ext_o-H2O}      
Band & $\lambda$ & E$_{\rm up}$ & Transition & Blend & Central $\lambda$ of fit & Flux & Error & FWHM & PACS FWHM & Ratio \\
& [$\mu$m] & [K] & $\nu$, $J_{Ka,Kc}-J_{Ka,Kc}$ & - & [$\mu$m] & [W\,m$^{-2}$] & [W\,m$^{-2}$] & [$\mu$m] & [$\mu$m] & - \\[2pt]
\hline\\[-8pt]
\endfirsthead
\hline
\multicolumn{10}{r}{Table continues in the next page.}
\endfoot
\caption{continued.}\\
Band & $\lambda$ & E$_{\rm up}$ & Transition & Blend & Central $\lambda$ of fit & Flux & Error & FWHM & PACS FWHM & Ratio \\
& [$\mu$m] & [K] & $\nu$, $J_{Ka,Kc}-J_{Ka,Kc}$ & - & [$\mu$m] & [W\,m$^{-2}$] & [W\,m$^{-2}$] & [$\mu$m] & [$\mu$m] & - \\[2pt]
\hline\\[-8pt]
\endhead
\hline\\[-8pt]
\endlastfoot
B2A & 56.816 & 1324.0 & $\nu =0, 9_{0,9}-8_{1,8}$ & No & 56.814 & 1.48e-15 & 4.0e-16 & 0.040 & 0.039 & 1.034\\
&57.268 & 3614.8 & $\nu =1, 9_{0,9}-8_{1,8}$ & No &  57.271 & 6.51e-16 & 3.0e-16 & 0.041 & 0.039 & 1.057\\
& 57.684 & 5853.5 & $\nu =2, 9_{1,9}-8_{0,8}$ &  Yes$^{\rm\, p}$ & 57.660 & 6.76e-15 & 1.4e-15 & 0.063 & 0.039 & 1.602\\
& 58.699 & 550.4 & $\nu =0, 4_{3,2}-3_{2,1}$ & No &  58.706 & 1.96e-15 & 4.1e-16 & 0.040 & 0.039 & 1.029\\
& 60.492 & 2744.8 & $\nu =1, 3_{3,0}-2_{2,1}$ & No &  60.493 & 4.93e-16 & 1.7e-16 & 0.026 & 0.039 & 0.650\\
& 62.335 & 3109.8 & $\nu =1, 6_{2,5}-5_{1,4}$ & No &  62.340 & 4.55e-16 & 1.7e-16 & 0.043 & 0.039 & 1.085\\
& 62.397 & 3673.2 & $\nu =1, 6_{5,2}-7_{2,5}$ &  Yes$^{\rm\, o,p}$ & 62.426 & 5.29e-16 & 1.9e-16 & 0.029 & 0.039 & 0.748\\
& 62.418 & 1845.9 & $\nu =0, 9_{3,6}-8_{4,5}$ &  Yes$^{\rm\, o,p}$ & 62.426 & 5.29e-16 & 1.9e-16 & 0.029 & 0.039 & 0.748\\
& 62.928 & 1552.6 & $\nu =0, 9_{1,8}-9_{0,9}$ & No &  62.930 & 2.86e-16 & 1.3e-16 & 0.023 & 0.039 & 0.583\\
& 63.324 & 1070.7 & $\nu =0, 8_{1,8}-7_{0,7}$ & No &  63.320 & 1.72e-15 & 3.8e-16 & 0.044 & 0.039 & 1.107\\
& 63.685 & 3363.5 & $\nu =1, 8_{1,8}-7_{0,7}$ & No &  63.690 & 3.77e-16 & 1.6e-16 & 0.038 & 0.039 & 0.974\\
& 63.914 & 1503.7 & $\nu =0, 6_{6,1}-6_{5,2}$ &  Yes$^{\rm\, o,p}$ & 63.940 & 1.38e-15 & 4.7e-16 & 0.057 & 0.039 & 1.458\\
& 63.955 & 1749.9 & $\nu =7, 6_{1,0}-5_{2,0}$ &  Yes$^{\rm\, o,p}$ & 63.940 & 1.38e-15 & 4.7e-16 & 0.057 & 0.039 & 1.458\\
& 65.166 & 795.5 & $\nu =0, 6_{2,5}-5_{1,4}$ & No &  65.172 & 1.55e-15 & 3.3e-16 & 0.038 & 0.039 & 0.957\\
& 66.093 & 1013.2 & $\nu =0, 7_{1,6}-6_{2,5}$ & No &  66.101 & 1.18e-15 & 2.7e-16 & 0.039 & 0.039 & 0.999\\
& 66.438 & 410.7 & $\nu =0, 3_{3,0}-2_{2,1}$ & No &  66.440 & 2.07e-15 & 4.3e-16 & 0.036 & 0.039 & 0.915\\
& 67.269 & 519.1 & $\nu =0, 3_{3,0}-3_{0,3}$ &  Yes & 67.272 & 2.26e-15 & 4.7e-16 & 0.052 & 0.039 & 1.322\\
& 67.365 & 3323.3 & $\nu =1, 7_{1,6}-6_{2,5}$ &  Yes & 67.373 & 5.89e-16 & 1.7e-16 & 0.055 & 0.039 & 1.391\\
& 70.287 & 2617.7 & $\nu =1, 3_{2,1}-2_{1,2}$ & No &  70.287 & 8.53e-16 & 2.5e-16 & 0.031 & 0.039 & 0.799\\
& 70.703 & 1274.2 & $\nu =0, 8_{2,7}-8_{1,8}$ & No &  70.702 & 8.57e-16 & 2.0e-16 & 0.042 & 0.039 & 1.065\\
& 71.947 & 843.5 & $\nu =0, 7_{0,7}-6_{1,6}$ & No &  71.956 & 1.60e-15 & 3.4e-16 & 0.039 & 0.039 & 0.986\\
\hline\\[-8pt]
B2B & 72.522 & 3137.6 & $\nu =1, 7_{0,7}-6_{1,6}$ & No & 72.543 & 6.40e-16 & 1.6e-16 & 0.036 & 0.039 & 0.924\\
& 73.415 & 5800.0 & $\nu =2, 8_{1,7}-8_{0,8}$ & No$^{\rm m}$ &  73.431 & 2.02e-15 & 4.2e-16 & 0.029 & 0.039 & 0.743\\
& 73.745 & 2745.3 & $\nu =1, 4_{2,3}-3_{1,2}$ & No &  73.763 & 6.24e-16 & 1.8e-16 & 0.040 & 0.039 & 1.021\\
& 74.945 & 1125.8 & $\nu =0, 7_{2,5}-6_{3,4}$ &  Yes & 74.966 & 1.64e-15 & 3.6e-16 & 0.054 & 0.039 & 1.374\\
& 75.381 & 305.3 & $\nu =0, 3_{2,1}-2_{1,2}$ & No &  75.407 & 3.51e-15 & 7.1e-16 & 0.031 & 0.039 & 0.783\\
& 75.830 & 1278.6 & $\nu =0, 6_{5,2}-6_{4,3}$ &  Yes$^{\rm\, p}$ & 75.847 & 1.01e-15 & 2.7e-16 & 0.063 & 0.039 & 1.621\\
& 75.910 & 1067.7 & $\nu =0, 5_{5,0}-5_{4,1}$ & No &  75.923 & 5.62e-16 & 1.7e-16 & 0.031 & 0.039 & 0.805\\
& 77.761 & 1524.9 & $\nu =0, 7_{5,2}-7_{4,3}$ & No &  77.785 & 2.91e-16 & 8.1e-17 & 0.032 & 0.039 & 0.827\\
& 78.742 & 432.2 & $\nu =0, 4_{2,3}-3_{1,2}$ & No &  78.766 & 2.97e-15 & 6.0e-16 & 0.039 & 0.039 & 1.019\\
& 78.946 & 3450.9 & $\nu =1, 6_{4,3}-6_{3,4}$ & Yes$^{\rm\, p}$ &  78.950 & 1.41e-15 & 3.0e-16 & 0.042 & 0.039 & 1.078\\
& 79.819 & 5358.0 & $\nu =2, 6_{1,5}-5_{2,4}$ &  Yes & 79.833 & 4.01e-16 & 1.3e-16 & 0.054 & 0.039 & 1.398\\
& 80.139 & 3064.2 & $\nu =1, 4_{4,1}-4_{3,2}$ &  Yes & 80.157 & 6.18e-16 & 1.6e-16 & 0.051 & 0.038 & 1.323\\
& 81.405 & 1729.4 & $\nu =0, 9_{2,7}-9_{1,8}$ & No &  81.425 & 3.77e-16 & 1.1e-16 & 0.039 & 0.038 & 1.027\\
& 82.031 & 643.5 & $\nu =0, 6_{1,6}-5_{0,5}$ & No &  82.052 & 1.98e-15 & 4.1e-16 & 0.032 & 0.038 & 0.846\\
& 82.726 & 3442.6 & $\nu =1, 7_{2,5}-6_{3,4}$ & No &  82.757 & 3.51e-16 & 1.2e-16 & 0.042 & 0.038 & 1.116\\
& 82.977 & 1447.6 & $\nu =0, 8_{3,6}-8_{2,7}$ & No &  82.998 & 2.58e-16 & 9.2e-17 & 0.025 & 0.038 & 0.664\\
& 83.724 & 5552.5 & $\nu =2, 7_{2,6}-7_{1,7}$ &  Yes & 83.737 & 5.00e-16 & 1.6e-16 & 0.064 & 0.038 & 1.690\\
& 85.769 & 1615.4 & $\nu =0, 8_{4,5}-8_{3,6}$ &  Yes$^{\rm\, p}$ &  85.796 & 3.66e-16 & 8.9e-17 & 0.030 & 0.037 & 0.803\\
& 92.811 & 1088.8 & $\nu =0, 6_{4,3}-6_{3,4}$ & No &  92.826 & 4.43e-16 & 9.5e-17 & 0.034 & 0.035 & 0.967\\
& 93.214 & 4838.3 & $\nu =2, 3_{2,2}-2_{1,1}$ & No &  93.233 & 2.87e-16 & 6.6e-17 & 0.033 & 0.035 & 0.936\\
& 94.644 & 795.5 & $\nu =0, 6_{2,5}-6_{1,6}$ & No &  94.665 & 5.51e-16 & 1.2e-16 & 0.029 & 0.034 & 0.829\\
& 94.705 & 702.3 & $\nu =0, 4_{4,1}-4_{3,2}$ & No &  94.725 & 6.39e-16 & 1.4e-16 & 0.035 & 0.034 & 1.009\\
& 95.176 & 1957.2 & $\nu =0, 9_{4,5}-8_{5,4}$ & No &  95.193 & 2.78e-16 & 7.6e-17 & 0.029 & 0.034 & 0.854\\
& 97.785 & 5003.7 & $\nu =2, 5_{1,5}-4_{0,4}$ &  Yes$^{\rm\, p}$ &  97.804 & 1.50e-16 & 4.2e-17 & 0.021 & 0.033 & 0.628\\
& 98.232 & 2506.9 & $\nu =1, 2_{2,1}-1_{1,0}$ & No &  98.254 & 2.87e-16 & 6.9e-17 & 0.036 & 0.033 & 1.103\\
& 98.494 & 878.2 & $\nu =0, 5_{4,1}-5_{3,2}$ & No &  98.511 & 3.05e-16 & 7.0e-17 & 0.029 & 0.033 & 0.880\\
\hline\\[-8pt]
R1A & 104.094 & 933.8 & $\nu =0, 6_{3,4}-6_{2,5}$ & No &  104.098 & 1.39e-16 & 5.8e-17 & 0.064 & 0.111 & 0.575\\
& 107.704 & 2878.9 & $\nu =1, 5_{0,5}-3_{3,0}$ & No &  107.745 & 1.74e-16 & 6.8e-17 & 0.094 & 0.113 & 0.838\\
& 108.073 & 194.1 & $\nu =0, 2_{2,1}-1_{1,0}$ & No &  108.100 & 1.69e-15 & 3.4e-16 & 0.112 & 0.113 & 0.992\\
& 111.483 & 2621.0 & $\nu =1, 4_{1,4}-3_{0,3}$ &  Yes$^{\rm\, p}$ & 111.586 & 3.13e-16 & 8.6e-17 & 0.163 & 0.114 & 1.429\\
& 112.511 & 1339.9 & $\nu =0, 7_{4,3}-7_{3,4}$ & No &  112.550 & 1.53e-16 & 5.4e-17 & 0.099 & 0.115 & 0.859\\
& 113.537 & 323.5 & $\nu =0, 4_{1,4}-3_{0,3}$ & No &  113.538 & 1.51e-15 & 3.1e-16 & 0.119 & 0.115 & 1.036\\
& 116.779 & 1212.0 & $\nu =0, 7_{3,4}-6_{4,3}$ & No &  116.789 & 5.88e-16 & 1.4e-16 & 0.141 & 0.116 & 1.207\\
& 121.722 & 550.4 & $\nu =0, 4_{3,2}-4_{2,3}$ & No &  121.732 & 3.57e-16 & 8.3e-17 & 0.104 & 0.118 & 0.882\\
& 127.884 & 1125.8 & $\nu =0, 7_{2,5}-7_{1,6}$ & No &  127.907 & 1.24e-16 & 4.4e-17 & 0.091 & 0.120 & 0.757\\
& 132.408 & 432.2 & $\nu =0, 4_{2,3}-4_{1,4}$ & No &  132.453 & 5.17e-16 & 1.1e-16 & 0.116 & 0.122 & 0.950\\
& 133.549 & 1447.6 & $\nu =0, 8_{3,6}-7_{4,3}$ & No$^{\rm m}$ & 133.563 & 3.40e-16 & 8.1e-17 & 0.101 & 0.122 & 0.830\\
& 134.935 & 574.8 & $\nu =0, 5_{1,4}-5_{0,5}$ & No &  134.966 & 3.08e-16 & 7.4e-17 & 0.140 & 0.123 & 1.142\\
& 136.496 & 410.7 & $\nu =0, 3_{3,0}-3_{2,1}$ & No &  136.513 & 4.46e-16 & 1.0e-16 & 0.147 & 0.123 & 1.196\\
\hline
R1B & 156.265 & 642.5 & $\nu =0, 5_{2,3}-4_{3,2}$ &  Yes$^{\rm\, p}$ & 156.262 & 1.25e-15 & 2.5e-16 & 0.156 & 0.126 & 1.240\\
& 159.051 & 1615.4 & $\nu =0, 8_{4,5}-7_{5,2}$ & No$^{\rm m}$ &  159.090 & 3.16e-16 & 7.0e-17 & 0.094 & 0.126 & 0.744\\
& 160.510 & 732.1 & $\nu =0, 5_{3,2}-5_{2,3}$ & No &  160.531 & 2.80e-16 & 6.4e-17 & 0.107 & 0.126 & 0.853\\
& 166.815 & 1212.0 & $\nu =0, 7_{3,4}-7_{2,5}$ &  Yes$^{\rm\, o}$ & 166.823 & 1.70e-16 & 4.3e-17 & 0.148 & 0.125 & 1.181\\
& 166.827 & 5428.8 & $\nu =2, 6_{2,4}-6_{1,5}$ &  Yes$^{\rm\, o}$ & 166.823 & 1.70e-16 & 4.3e-17 & 0.148 & 0.125 & 1.181\\
& 170.928 & 2413.0 & $\nu =1, 2_{1,2}-1_{0,1}$ & No &  170.957 & 1.27e-16 & 3.5e-17 & 0.140 & 0.125 & 1.123\\
& 174.626 & 196.8 & $\nu =0, 3_{0,3}-2_{1,2}$ & No$^{\rm m}$ &  174.641 & 1.29e-15 & 2.6e-16 & 0.121 & 0.124 & 0.974\\
& 179.527 & 114.4 & $\nu =0, 2_{1,2}-1_{0,1}$ & No &  179.553 & 1.11e-15 & 2.2e-16 & 0.100 & 0.122 & 0.820\\
& 180.488 & 194.1 & $\nu =0, 2_{2,1}-2_{1,2}$ & No &  180.514 & 4.41e-16 & 9.2e-17 & 0.115 & 0.122 & 0.942\\
\end{longtable}
}
\end{center}

\begin{table*}[hb!]
\caption{Extracted p-H$_2^{16}$O lines from the PACS observations. The blended and masering lines are given as in Table \ref{table:ext_o-H2O}.}             
\label{table:ext_p-H2O}      
\centering
\begin{tabular}{l @{}c c c l c c c c c c c } 
Band & $\lambda$ & E$_{\rm up}$ & Transition & Blend & Central $\lambda$ of fit & Flux & Error & FWHM & PACS FWHM & Ratio \\
& [$\mu$m] & [K] & $\nu$, $J_{Ka,Kc}-J_{Ka,Kc}$ & - & [$\mu$m] & [W\,m$^{-2}$] & [W\,m$^{-2}$] & [$\mu$m] & [$\mu$m] & - \\[2pt]
\hline\\[-8pt]
B2A & 56.325 & 1048.5 & $\nu =0, 4_{3,1}-3_{2,2}$ & No & 56.327 & 1.80e-15 & 3.8e-16 & 0.036 & 0.039 & 0.921\\
&59.987 & 1708.9 & $\nu =0, 7_{2,6}-6_{1,5}$ &  Yes & 59.994 & 1.75e-15 & 4.1e-16 & 0.060 & 0.039 & 1.539\\
&60.162 & 2273.8 & $\nu =0, 8_{2,6}-7_{3,5}$ &  Yes & 60.189 & 1.51e-15 & 4.5e-16 & 0.109 & 0.039 & 2.774\\
& 60.989 & 2631.5 & $\nu =1, 3_{3,1}-2_{2,0}$ & No & 60.985 & 7.21e-16 & 1.9e-16 & 0.037 & 0.039 & 0.934\\
& 61.809 & 552.3 & $\nu =0, 4_{3,1}-4_{0,4}$ & No & 61.801 & 1.05e-15 & 2.4e-16 & 0.035 & 0.039 & 0.886\\
&62.432 & 1554.5 & $\nu =0, 9_{2,8}-9_{1,9}$ &  Yes$^{\rm\, o}$ & 62.447 & 1.28e-15 & 3.6e-16 & 0.094 & 0.039 & 2.392\\
& 63.458 & 1070.6 & $\nu =0, 8_{0,8}-7_{1,7}$ & No & 63.468 & 7.12e-16 & 2.0e-16 & 0.040 & 0.039 & 1.005\\
&63.928 & 1503.7 & $\nu =0, 6_{6,0}-6_{5,1}$ &  Yes$^{\rm\, o,p}$ & 63.940 & 1.36e-15 & 4.6e-16 & 0.057 & 0.039 & 1.439\\
&63.949 & 3363.2 & $\nu =1, 8_{0,8}-7_{1,7}$ &  Yes$^{\rm\, o,p}$ & 63.940 & 1.36e-15 & 4.6e-16 & 0.057 & 0.039 & 1.439\\
&67.089 & 410.4 & $\nu =0, 3_{3,1}-2_{2,0}$ &  Yes & 67.094 & 2.05e-15 & 4.2e-16 & 0.054 & 0.039 & 1.370\\
& 71.067 & 598.9 & $\nu =0, 5_{2,4}-4_{1,3}$ & No & 71.062 & 1.20e-15 & 2.5e-16 & 0.037 & 0.039 & 0.949\\
\hline\\[-8pt]
B2B & 71.540 & 843.8 & $\nu =0, 7_{1,7}-6_{0,6}$ & No & 71.561 & 9.60e-16 & 2.1e-16 & 0.033 & 0.039 & 0.839\\
& 71.783 & 3138.2 & $\nu =1, 7_{1,7}-6_{0,6}$ &  Yes$^{\rm\, p}$ & 71.807 & 5.17e-16 & 1.5e-16 & 0.040 & 0.039 & 1.028\\
& 71.788 & 1067.7 & $\nu =0, 5_{5,1}-6_{2,4}$ &  Yes$^{\rm\, p}$ & 71.807 & 5.17e-16 & 1.5e-16 & 0.040 & 0.039 & 1.028\\
& 78.928 & 781.2 & $\nu =0, 6_{1,5}-5_{2,4}$ & No & 78.951 & 1.45e-15 & 3.1e-16 & 0.042 & 0.039 & 1.096\\
& 80.222 & 1929.3 & $\nu =0, 9_{4,6}-9_{3,7}$ & No & 80.238 & 3.36e-16 & 1.1e-16 & 0.042 & 0.038 & 1.096\\
& 80.557 & 1807.1 & $\nu =0, 8_{5,3}-8_{4,4}$ & No & 80.587 & 2.43e-16 & 1.0e-16 & 0.050 & 0.038 & 1.314\\
& 81.216 & 1021.0 & $\nu =0, 7_{2,6}-7_{1,7}$ & No & 81.223 & 5.97e-16 & 1.4e-16 & 0.044 & 0.038 & 1.148\\
&81.690 & 1511.0 & $\nu =0, 8_{3,5}-7_{4,4}$ &  Yes & 81.726 & 5.78e-16 & 1.7e-16 & 0.051 & 0.038 & 1.329\\
&81.893 & 3088.1 & $\nu =1, 6_{1,5}-5_{2,4}$ &  Yes & 81.919 & 5.51e-16 & 1.7e-16 & 0.055 & 0.038 & 1.428\\
& 83.284 & 642.7 & $\nu =0, 6_{0,6}-5_{1,5}$ & No & 83.295 & 1.66e-15 & 3.4e-16 & 0.036 & 0.038 & 0.949\\
& 84.068 & 2937.8 & $\nu =1, 6_{0,6}-5_{1,5}$ & No & 84.086 & 4.68e-16 & 1.1e-16 & 0.044 & 0.038 & 1.154\\
& 85.781 & 3452.0 & $\nu =1, 6_{4,2}-6_{3,3}$ & Yes$^{\rm\, o}$ & 85.796 & 4.10e-16 & 9.4e-17 & 0.033 & 0.037 & 0.872\\
& 89.988 & 296.8 & $\nu =0, 3_{2,2}-2_{1,1}$ & No & 89.998 & 2.08e-15 & 4.4e-16 & 0.040 & 0.036 & 1.109\\
& 92.150 & 2508.6 & $\nu =1, 2_{2,0}-1_{1,1}$ & No & 92.175 & 4.84e-16 & 1.0e-16 & 0.035 & 0.035 & 0.998\\
&93.383 & 1175.1 & $\nu =0, 7_{3,5}-7_{2,6}$ &  Yes & 93.380 & 7.13e-16 & 1.5e-16 & 0.059 & 0.035 & 1.681\\
&94.210 & 877.9 & $\nu =0, 5_{4,2}-5_{3,3}$ &  Yes$^{\rm\, o}$ & 94.222 & 5.34e-16 & 1.2e-16 & 0.055 & 0.035 & 1.587\\
&94.897 & 2766.7 & $\nu =1, 5_{1,5}-4_{0,4}$ &  Yes & 94.899 & 6.06e-16 & 1.3e-16 & 0.071 & 0.034 & 2.057\\
&95.627 & 470.0 & $\nu =0, 5_{1,5}-4_{0,4}$ &  Yes & 95.643 & 1.97e-15 & 4.0e-16 & 0.050 & 0.034 & 1.454\\
\hline\\[-8pt]
R1A & 111.628 & 598.9 & $\nu =0, 5_{2,4}-5_{1,5}$ & No & 111.617 & 2.29e-16 & 8.2e-17 & 0.097 & 0.114 & 0.853\\
& 113.948 & 725.1 & $\nu =0, 5_{3,3}-5_{2,4}$ & No & 113.962 & 3.16e-16 & 8.1e-17 & 0.120 & 0.115 & 1.037\\
& 125.354 & 319.5 & $\nu =0, 4_{0,4}-3_{1,3}$ & No & 125.383 & 1.03e-15 & 2.1e-16 & 0.126 & 0.120 & 1.054\\
&126.714 & 410.4 & $\nu =0, 3_{3,1}-3_{2,2}$ &  Yes & 126.689 & 3.22e-16 & 8.5e-17 & 0.206 & 0.120 & 1.717\\
&128.259 & 2615.0 & $\nu =1, 4_{0,4}-3_{1,3}$ &  Yes & 128.284 & 2.59e-16 & 7.3e-17 & 0.192 & 0.121 & 1.594\\
& 138.528 & 204.7 & $\nu =0, 3_{1,3}-2_{0,2}$ & No$^{\rm\, m}$ & 138.549 & 1.02e-15 & 2.1e-16 & 0.125 & 0.123 & 1.010\\
\hline\\[-8pt]
R1B & 144.518 & 396.4 & $\nu =0, 4_{1,3}-3_{2,2}$ & No & 144.555 & 7.65e-16 & 1.6e-16 & 0.109 & 0.125 & 0.876\\
& 146.923 & 552.3 & $\nu =0, 4_{3,1}-4_{2,2}$ & No & 146.946 & 2.16e-16 & 4.8e-17 & 0.106 & 0.125 & 0.851\\
&156.194 & 296.8 & $\nu =0, 3_{2,2}-3_{1,3}$ &  Yes$^{\rm\, o}$ & 156.262 & 1.25e-15 & 2.5e-16 & 0.152 & 0.126 & 1.210\\
& 167.035 & 867.3 & $\nu =0, 6_{2,4}-6_{1,5}$ & No & 167.063 & 7.87e-17 & 2.5e-17 & 0.091 & 0.125 & 0.727\\
& 169.739 & 1175.1 & $\nu =0, 7_{3,5}-6_{4,2}$ & No & 169.764 & 2.14e-16 & 4.7e-17 & 0.101 & 0.125 & 0.806\\
& 170.139 & 951.9 & $\nu =0, 6_{3,3}-6_{2,4}$ & No & 170.160 & 1.67e-16 & 4.5e-17 & 0.125 & 0.125 & 1.001\\
& 187.111 & 396.4 & $\nu =0, 4_{1,3}-4_{0,4}$ & No & 187.124 & 1.58e-16 & 3.4e-17 & 0.086 & 0.119 & 0.721\\
\hline
\end{tabular}
\end{table*}

\end{appendix}

\end{document}